\documentclass[sn-apa]{sn-jnl}

\usepackage{graphicx}%
\usepackage{multirow}%
\usepackage{amsmath,amssymb,amsfonts}%
\usepackage{amsthm}%
\usepackage{mathrsfs}%
\usepackage[title]{appendix}%
\usepackage{xcolor}%
\usepackage{textcomp}%
\usepackage{manyfoot}%
\usepackage{booktabs}%
\usepackage{algorithm}%
\usepackage{algorithmicx}%
\usepackage{algpseudocode}%
\usepackage{listings}%

\usepackage{multirow}
\usepackage{multicol}
\usepackage{makecell}
\usepackage[most]{tcolorbox}
\usepackage{array}
\usepackage{dcolumn}
\usepackage{pifont}
\usepackage{dirtree}
\usepackage{forest}
\usepackage{pdflscape}
\usepackage{afterpage}
\usepackage{float}
\usepackage{enumerate}
\usepackage{longtable}

\newcolumntype{d}[1]{D{#1}{}{-1}}

\newcommand*{\img}[1]{%
	\raisebox{-.3\baselineskip}{%
		\includegraphics[
		height=\baselineskip,
		width=\baselineskip,
		keepaspectratio,
		]{#1}%
	}%
}

\newcolumntype{C}[1]{>{\centering\arraybackslash}p{#1}}

\theoremstyle{thmstyleone}%
\theoremstyle{thmstyletwo}%

\theoremstyle{thmstylethree}%

\begin{document}

\title[Benchmarking LLMs on Homework Assessment in Circuit Analysis]{Benchmarking Large Language Models on Homework Assessment in Circuit Analysis}

\author{\fnm{Liangliang} \sur{Chen}}
\author{\fnm{Zhihao} \sur{Qin}}
\author{\fnm{Yiming} \sur{Guo}}
\author{\fnm{Jacqueline} \sur{Rohde}}
\author*{\fnm{Ying} \sur{Zhang*}}\email{yzhang@gatech.edu}

\affil{\orgdiv{School of Electrical and Computer Engineering}, \orgname{Georgia Institute of Technology}, \orgaddress{\city{Atlanta},  \state{GA} \postcode{30332}, \country{USA}}}

\abstract{Large language models (LLMs) have the potential to revolutionize various fields, including code development, robotics, finance, and education, due to their extensive prior knowledge and rapid advancements. This paper investigates how LLMs can be leveraged in engineering education. Specifically, we benchmark the capabilities of different LLMs, including GPT-3.5 Turbo, GPT-4o, and Llama 3 70B, in assessing homework for an undergraduate-level circuit analysis course. We have developed a novel dataset consisting of official reference solutions and real student solutions to problems from various topics in circuit analysis. To overcome the limitations of image recognition in current state-of-the-art LLMs, the solutions in the dataset are converted to LaTeX format. Using this dataset, a prompt template is designed to test five metrics of student solutions: completeness, method, final answer, arithmetic error, and units. The results show that GPT-4o and Llama 3 70B perform significantly better than GPT-3.5 Turbo across all five metrics, with GPT-4o and Llama 3 70B each having distinct advantages in different evaluation aspects. Additionally, we present insights into the limitations of current LLMs in several aspects of circuit analysis. Given the paramount importance of ensuring reliability in LLM-generated homework assessment to avoid misleading students, our results establish benchmarks and offer valuable insights for the development of a reliable, personalized tutor for circuit analysis---a focus of our future work. Furthermore, the proposed evaluation methods can be generalized to a broader range of courses for engineering education in the future.}




\maketitle

\setlength{\arraycolsep}{1.5pt}

\section{Introduction}
The rapid development of large language models (LLMs) \citep{brown2020language, ouyang2022training, achiam2023gpt, touvron2023llama, touvron2023llama2, team2023gemini, reid2024gemini} is revolutionizing many fields, including software engineering \citep{ozkaya2023application, ebert2023generative}, robotics \citep{xie2023text2reward, ma2023eureka, chen2024rlingua}, and education \citep{jeon2023large, ansari2023mapping, kasneci2023chatgpt}. With substantial prior knowledge, LLMs can generate reasonable and reliable responses in various domains through appropriate prompt designs \citep{min2022rethinking, liu2023pre}. In the field of LLM-assisted education, existing works have demonstrated the usefulness of LLMs in programming education and coding understanding \citep{hellas2023exploring, nam2024using}, medical education \citep{safranek2023role}, mechanical engineering \citep{mustapha2024bard}, and chemical engineering education \citep{tsai2023exploring}, among others. It is important to note that the application of LLMs in educational fields requires careful evaluation of the models' capacities, as the inevitable hallucinations of LLMs \citep{huang2023survey, xu2024hallucination} may result in incorrect responses that mislead students. In this paper, we investigate how LLMs can be used in an undergraduate-level circuit analysis course, with the long-term aim of building a reliable and intelligent personalized virtual tutor for students majoring in electrical engineering \citep{rohde2024predictors}.

The learning materials for circuit analysis encompass various formats of information, including text, equations, and circuit diagrams \citep{svoboda2013introduction}. A primary challenge for leveraging LLMs in a circuit analysis course lies in their ability to reliably recognize and interpret this diverse information. Our experiment reveals that even state-of-the-art vision-language models (VLMs) and multimodal language models struggle to reliably interpret basic circuit diagrams, rendering them incapable of independently solving circuit analysis problems without external references. In this paper, we highlight one specific area where state-of-the-art LLMs can contribute to circuit analysis education: homework assessment. In this context, the reference solution typically contains sufficient information to address most problems. Therefore, we include the reference solution directly in the prompt, bypassing the need for LLMs to interpret circuit diagrams. To support this study, we developed a novel dataset comprising real student solutions in LaTeX format and used it to benchmark three LLMs, evaluating their capabilities in homework assessment for circuit analysis. \textit{To the best of our knowledge, this is the first exploration of how LLMs can assist in undergraduate-level circuit analysis education.}

The main contributions of this paper are summarized as follows. 
\begin{itemize}
\item [i)] We construct a novel dataset that consists of 283 students' solutions from 119 problems on various topics in an undergraduate circuit analysis course. To facilitate the evaluation of current state-of-the-art LLMs, which are limited in their ability to recognize handwritten solutions, we handcrafted the dataset by converting both the official reference solutions and students' solutions to the LaTeX format using the Mathpix snipping tool. This dataset reflects the students' real homework completion behaviors in circuit analysis and serves as a reliable data source for evaluating the homework assessment capabilities of different LLMs.

\item [ii)] We test the homework assessment capabilities of three LLMs (GPT-3.5 Turbo, GPT-4o, and Llama 3 70B) using our dataset. In a designed prompt template, we ask the LLMs to assess student's homework solutions based on five different metrics: completeness, method, final answer, arithmetic error, and units. These metrics cover the main aspects in which students are prone to make mistakes in circuit analysis homework. We provide detailed results and systematic analyses of the performance of the considered LLMs in these metrics. We focus our analysis on the response characteristics of GPT-4o and Llama 3 70B, the better-performing options of the three models. This focus provides insights into the models' specific strengths and weaknesses in assessing homework for an undergraduate-level circuit analysis course. The evaluation framework proposed in this paper can also be generalized to a broader range of courses in engineering education.

\item [iii)] We discuss the limitations of current state-of-the-art LLMs in this context, including circuit diagram recognition, handwritten and typed solution recognition, and mathematical calculations. These identified limitations not only provide insights into strategies to navigate the limits of currently available LLMs for homework assessment but also suggest several possible improvement directions to build LLMs specific to the circuit analysis course, as well as a broader field of engineering education.
\end{itemize}

\section{Related Work}
\label{S2}
This section provides the related work in the aspects of large language model evaluations, large language models for education, and automatic homework assessment. 

\textbf{Large Language Model Evaluations:} LLMs, as general-purpose language generation tools, have experienced rapid development in recent years, including advancements such as GPT-3 \citep{brown2020language}, Llama \citep{touvron2023llama}, Llama 2 \citep{touvron2023llama2}, and multimodal models like GPT-4 \citep{achiam2023gpt}, Gemini \citep{team2023gemini}, and Llama 3. In addition to technical reports released by model developers, researchers from various fields have evaluated the domain-specific performances of different LLMs. For instance, \cite{frieder2023mathematical} measured the mathematical capabilities of ChatGPT and GPT-4 using their collected datasets. Similarly, \cite{guo2023can} provided a comprehensive benchmark for five LLMs across eight chemistry tasks, and \cite{du2024mercury} assessed the computational efficiency of code LLMs. Other fields where LLMs have been evaluated include scientific literature analysis \citep{cai2024sciassess}, structured table data understanding \citep{sui2024table}, legal reasoning \citep{guha2024legalbench}, finance \citep{xie2024finben}, automated planning \citep{valmeekam2024planbench}, etc. With the long-term goal of creating a reliable personalized virtual tutor for students majoring in electrical engineering, we benchmark LLMs in this paper on homework assessment for an undergraduate-level circuit analysis course. 

\textbf{Large Language Models for Education:} 
Due to their vast internal knowledge, LLMs have the potential to provide personalized support for students in various domains \citep{kasneci2023chatgpt, jeon2023large}. Since the reliability of responses is essential in the field of education to avoid misleading students, many existing works have evaluated LLMs' capabilities across different disciplines, including mathematics \citep{frieder2023mathematical}, medicine \citep{thirunavukarasu2023large, liu2024benchmarking}, language learning \citep{baskara2023exploring}, business \citep{cribben2023benefits}, computational social science \citep{ziems2024can}, etc. Specifically, in the fields of engineering education \citep{qadir2023engineering}, LLM benchmarks have been established from different points of view in mechanical engineering \citep{tian2024assessing}, chemical engineering \citep{tsai2023exploring}, software engineering \citep{meissner2024evalquiz}, and control engineering \citep{kevian2024capabilities}. Notably, \citet{ngoc2023ai} evaluated the efficacy of ChatGPT in an integrated circuit systems course. The results indicated that ChatGPT can generate good responses for theoretical and fact-based questions but fails to produce reliable responses for questions whose answers include graphics. The dataset in \citep{ngoc2023ai} consists of only 24 questions, which might introduce biased results due to the small size. In our paper, we present a dataset consisting of 283 student's solutions from 119 problems and use it to evaluate the homework assessment capabilities of various LLMs, including the recently released GPT-4o and Llama 3 70B.

\textbf{Automatic Homework Assessment:} Prior to the advent of LLMs, automatic assessment systems were widely designed based on semantic comparisons between official solutions and students' solutions \citep{lan2015mathematical, liu2019automatic, aldriye2019automated}. The semantic comparisons in traditional automatic grading systems require discipline-specific feature extraction, limiting the system's performance and generalization due to the variety of students' solutions. Recent works have explored the potential of LLMs in enhancing automatic assessment systems with their versatile semantic analysis ability and massive internal knowledge \citep{fagbohun2024beyond, lee2024applying, xie2024grade}. 

Two popular subjects for LLM-assessment systems are essay writing and mathematical problem-solving \citep{yancey2023rating, botelho2023leveraging, xiao2024automation, li2024automated, yoo2024dress}. Essay writing is a purely natural language task, making it well suited for LLM assessment. Although mathematics relies on a broader skillset, LLMs such as GPT-4 have demonstrated the capability to handle mathematical tasks up to the undergraduate level \citep{frieder2023mathematical}. In this paper, we evaluate LLMs' capabilities in homework assessment for circuit analysis, a fundamental course for undergraduate students majoring in electrical engineering. Compared with the existing benchmarks in other disciplines, circuit analysis is challenging because it involves circuit diagrams, mathematics, and domain-specific circuit knowledge.

\section{Methods}
\label{S3}
In this section, we illustrate the methods used in this paper, including the collection of the dataset and the evaluation of the LLMs.

\subsection{Dataset Collections}
\label{S31}
We collected data from an undergraduate-level circuit analysis course in the School of Electrical and Computer Engineering at the Georgia Institute of Technology. Specifically, the raw data includes the assignment problems from the textbook \citep{svoboda2013introduction}, the official reference solutions, and the handwritten student solutions from 37 students. In detail, we have 283 student's solutions in total from 119 problems. We categorize the data into six topics, with the relationships between the topics and problem indices in \cite{svoboda2013introduction} shown in Table \ref{TabA1}, and descriptions of the topics and problems presented in Table \ref{TabA0}. These topics comprehensively cover the contents of an undergraduate-level circuit analysis course. 

\begin{table}
	\centering
	\small
	\caption{\small Problem indexes and number of data in each topic}
	\label{TabA1}
	\begin{tabular}{ccc}
		\hline
		Topic & Problem Indexes & Number of Data\\
		\hline
		Electric Circuit Variables and Elements & 1.x-x, 2.x-x & 40 \\
		Analysis of Resistive Circuits & 3.x-x, 4.x-x, 5.x-x & 63 \\
		The Operational Amplifier & 6.x-x & 28 \\
		\makecell{The Complete Response of Circuits\\With Energy Storage Elements} & 7.x-x, 8.x-x, 9.x-x & 95 \\
		Sinusoidal Steady-State Analysis & 10.x-x & 29 \\
		Frequency Response & 13.x-x & 28 \\
		\hline
		\multicolumn{3}{l}{* ``x'' represents the index of either a section or a problem in a section.}
	\end{tabular}
\end{table}

\begin{small}
\begin{longtable}{ccc}
	\caption{Topic descriptions and problem descriptions}
	\label{TabA0}\\
	\hline
	Topic & Description & Problem Characteristics \\
	\hline
	\vspace{0.65cm} Var. \& Ele. & \makecell[{{p{5.0cm}}}]{Electric circuit variables and elements are basic concepts in circuit analysis. Students are expected to learn fundamental circuit concepts such as current, voltage, power, energy, and systems of units.} & \makecell[{{p{5.0cm}}}]{The problems in this topic typically require students to perform basic circuit calculations. For example, students may be asked to calculate a voltage value using Ohm's law. To provide homework assessments effectively, the LLMs being evaluated need to possess knowledge of basic circuit variables and elements.}\vspace{-0.7cm} \\ 
	\hline
	\vspace{1.10cm} Resist. Cir. & \makecell[{{p{5.0cm}}}]{The resistive circuit is one of the fundamental types of electrical circuits in which all circuit power is dissipated by resistors. A resistive circuit may consist of resistors and independent or dependent voltage and current sources, but not energy storage elements like capacitors and inductors. The key concepts that students need to learn in the analysis of resistive circuits include Kirchhoff's laws (KCL and KVL), series and parallel circuits, node voltage analysis, and mesh current analysis.} & \makecell[{{p{5.0cm}}}]{The students are typically asked to determine the values of resistances, currents, voltages, or powers of certain elements in given circuits. The final answers are typically numeric values with appropriate units in these cases. In order to provide reasonable homework assessments, the evaluated LLMs are required to understand the basic circuit analysis methods and be able to do some basic algebraic manipulations.} \vspace{-1.1cm} \\
	\hline
	\vspace{0.75cm} Op. Amp. & \makecell[{{p{5.0cm}}}]{The operational amplifier is a type of circuit element that can be used to build a circuit to perform mathematical operations. In our circuit analysis course, we mainly consider the ideal operational amplifier, which is characterized by having zero input currents and equal node voltages at its input terminals.} & \makecell[{{p{5.0cm}}}]{The homework problems in this topic typically require students to analyze circuits containing ideal operational amplifiers. The final answers may be in the form of current values, voltage values, or the gain of the circuit, which are numerical values with or without units. These problems test the LLM's ability to understand circuits with more advanced elements like the operational amplifier.} \vspace{-0.75cm} \\
	\hline
	\vspace{1.30cm} Com. Resp. & \makecell[{{p{5.0cm}}}]{The capacitor and the inductor are two basic energy storage elements. Since the constitutive equations of the capacitor and the inductor involve either differentiation or integration, differential equations are required to represent the circuit behaviors with these two energy storage elements. In the considered circuit analysis course, the key points in this topic include the basic properties, series and parallel connections of capacitors and inductors, the initial conditions in a switched circuit, and the complete responses of first-order or second-order circuits.} & \makecell[{{p{5.0cm}}}]{The problems in this topic involve more advanced devices and more complicated mathematical analyses. The currents or voltages in the circuits might be dynamic instead of static, meaning their values depend on time. The homework problems might ask for dynamic currents or voltages when $t>0$, differential equations for provided circuits, or inferences of device parameters (e.g., the capacitance of a certain capacitor) given circuit responses. To generate reasonable assessments, the LLMs need to understand different mathematical expressions, recognize differential equations, and perform higher-level mathematical calculations.} \vspace{-1.3cm} \\
	\hline
	\vspace{1.50cm} Sinusoidal & \makecell[{{p{5.0cm}}}]{The circuits considered in this topic are linear circuits with sinusoidal current or voltage sources. In this case, if there is only one sinusoidal source, the circuit response at its steady state will be a sine wave with the same frequency as the input. When there are multiple sources, we can use the principle of superposition to obtain the response. Students are expected to learn the behaviors of circuits that consist of one or more sinusoidal sources, the two representations (i.e., the sinusoid and the phasor) of a sinusoidal signal, the concept of impedance, and the use of basic circuit analysis methods (e.g., KCL and KVL) in the scenarios of sinusoidal steady-state analysis.} & \makecell[{{p{5.0cm}}}]{Most problems in this topic ask students to calculate the current or voltage responses of circuits with sinusoidal sources. Additionally, some problems may require students to infer the impedance of a particular component. To generate reasonable homework assessments, the tested LLMs need to understand not only the basic circuit principles and analysis methods but also the knowledge related to sinusoidal steady-state analysis, such as the conversions and calculations of sinusoidal time-domain values and phasors.} \vspace{-1.5cm} \\
	\hline
	\vspace{1.40cm} Freq. Resp. & \makecell[{{p{5.0cm}}}]{Based on the sinusoidal steady-state analysis, we can calculate the gain, phase shift, and network function of a linear circuit when there is a sinusoidal current or voltage input source. The frequency response is an equation that represents the gain and phase shift from the input to the output of a circuit. Apart from the formula representations, Bode plots are often used to visualize the frequency responses.} & \makecell[{{p{5.0cm}}}]{In our considered circuit analysis course, the problems about frequency responses typically ask students to either calculate the network function or draw a Bode plot for a given circuit. Students may also be required to infer some device parameters, such as the capacitance of a capacitor or the resistance of a resistor. The considered LLMs should be able to compare different network functions in terms of both their forms and coefficients and provide a reasonable assessment if the student's solution is incorrect. Due to the limited capacity of LLMs for image recognition, we do not ask the LLMs to assess Bode plots.} \vspace{-1.45cm} \\
	\hline
	\multicolumn{3}{l}{\makecell[{{p{12.55cm}}}]{* \ding{172} Var. \& Ele. = Electric Circuit Variables and Elements; \ding{173} Resist. Cir. = Analysis of Resistive Circuits; \ding{174} Op. Amp. = The Operational Amplifier; \ding{175} Com. Resp. = The Complete Response of Circuits With Energy Storage Elements; \ding{176} Sinusoidal = Sinusoidal Steady-State Analysis; \ding{177} Freq. Resp. = Frequency Response.}}
\end{longtable}
\end{small}

Due to the limitation that current state-of-the-art LLMs/VLMs cannot reliably recognize content in images such as text, mathematical equations, and circuit diagrams (see details in Section \ref{S5}), the typed official solutions and handwritten student solutions were converted to LaTeX format using the Mathpix snipping tool\footnote{See \url{https://mathpix.com}.}. The constructed dataset does not consist of circuit diagrams and problem statements. We also manually proofread and corrected the automatically generated LaTeX source files to eliminate typos. Additionally, we included a short summary of the final answer for each problem in the dataset to make it easy for the LLMs to determine if the final answer in a student's solution is correct. The evaluation results in Section \ref{S4} indicate that this method is effective for GPT-4o and Llama 3 70B to provide reasonable homework assessments. The detailed structure of the dataset is presented in Appendix \ref{SA}.

\subsection{LLM Evaluations}
\label{S32}

The prompt template for the evaluations of different LLMs is provided as follows. The components that need to be adapted to different problems and solutions are marked with a \colorbox{lightgray}{light gray} background. The official reference solution, along with the final answer, serves as a reference for LLMs to evaluate the student's solution and generate an assessment report.

\begin{center}
	\begin{tcolorbox}[breakable, enhanced, width=\textwidth, opacityfill=1., colback=white]
		Now you play the role of an instructor and need to provide feedback for a student's homework solution. This homework is from a course on circuit analysis. The official solution and student solution are provided in the LaTeX form as follows. The official solution is always correct and can serve as a benchmark for homework assessment. You can provide feedback based on the following aspects. Your feedback should be detailed and precise. \\
		
		1. Is the student's solution complete? In other words, does the student's solution answer the question? 
		
		2. Does the student use the correct method? 
		
		3. Are the student's final answers to the problem correct? 
		
		NOTE: The correct final answer(s): \colorbox{lightgray}{\{FINAL ANSWER GOES HERE\}}
		
		4. Is there any arithmetic error? Note that the student may use different variable notations from those in the standard solution. These notation differences only should not be regarded as errors, as long as the other parts are correct. 
		
		5. Are the units of all variables identified clearly and correctly throughout the calculation process? \\
		
		[Notes] 
		
		You need to consider the following aspects when giving feedback about the student's solution. 
		
		1. There might be some typos in the student's solutions. The student can be regarded as solving the problem correctly if all other steps except for the typo are correct. 
		
		2. Be careful to check the calculations. The numbers and their signs MUST be correct, and the errors of them cannot be regarded as typos. 
		
		3. The rounding errors during the calculation process should not be considered as calculation errors. 
		
		4. The equivalency between decimals and fractions should not be regarded as errors.
		
		5. If you think the student's solution is correct, you may just provide concise and brief assessments. When the student's solution is wrong, you need to provide detailed analyses about why the solution is wrong. \\
		
		[Official Solution] 
		
		\colorbox{lightgray}{\{OFFICIAL SOLUTION IN \LaTeX ~GOES HERE\}}\\
		
		[Student's Solution]
		
		\colorbox{lightgray}{\{STUDENT'S SOLUTION IN \LaTeX ~GOES HERE\}}

	\end{tcolorbox}
\end{center}

The prompt designed above assesses the student's homework in the following metrics, which cover the essential aspects of the homework assessment when the student's solutions were graded.\footnote{The student's solution were graded by the first author of this paper, Liangliang Chen, who was a teaching assistant of the circuit analysis course during the dataset collection period.}

\begin{itemize}
	\item [i)] {\textit{Completeness---The completeness of the student’s solution:}} This metric evaluates whether the student has completely answered all questions in the problems. 
	
	{\textit{Justification:}} Completing circuit analysis problems in full is the minimum requirement for students when they work on assigned homework. When LLMs are presented with both the official solution and the student's solution, applying completeness as a metric specifically prevents LLMs from marking a student's solution as correct if it is only partially complete and the completed portion is accurate. Note that this metric is not intended to assess the correctness of the final answer itself. A student's solution is considered correct as long as it reaches the final step of the solution and provides an answer to the problem.
	
	\item [ii)] {\textit{Method---The correctness of the method:}} This metric evaluates whether the student uses the correct method, regardless of arithmetic errors or typos.
	
	{\textit{Justification:}} A correct solution from a student often implies that they used a reasonable method. However, when a student's solution is incorrect, they may have used either an appropriate or an inappropriate method. Therefore, including a metric for the method in the prompt will help students determine whether they have adopted the correct approach. 
	
	\item [iii)] {\textit{Final Answer---The correctness of the final answers:}} This metric evaluates whether the final answers provided by the student are correct. 
	
	{\textit{Justification:}} The final answer requires the LLMs to compare the student’s final answer with that of the official reference solution. While the problem-solving process is also important, the final answer, typically brief, is straightforward for LLMs to assess. Including the assessment of final answers enhances the quality of LLM evaluations, as LLMs tend to provide consistent assessments over different metrics. For example, if a student's final answer does not match the correct answer, LLMs are likely to offer negative assessments, encouraging them to identify the student's incorrect steps.
	
	\item [iv)] {\textit{Arithmetic---Arithmetic error:}} This metric evaluates whether there is any arithmetic error in the student’s solution. 
	
	{\textit{Justification:}} Accurate arithmetic calculations are essential for deriving correct answers in circuit analysis problems. When a student’s solution contains an arithmetic error, identifying these errors can be very beneficial for learning. Therefore, we include arithmetic as one of the metrics to evaluate the LLMs' ability to provide assessments on any arithmetic errors in students' solutions.
	
	\item [v)] {\textit{Unit---Units of variables:}} This metric evaluates whether the students use appropriate units for the circuit variables in their solutions.
	
	{\textit{Justification:}} Correctly using units for circuit variables is important for students to master basic circuit concepts. Therefore, we include a metric to check whether units are appropriately used in students' solutions.
\end{itemize}

\begin{figure}
	\centering
	\includegraphics[width=\textwidth]{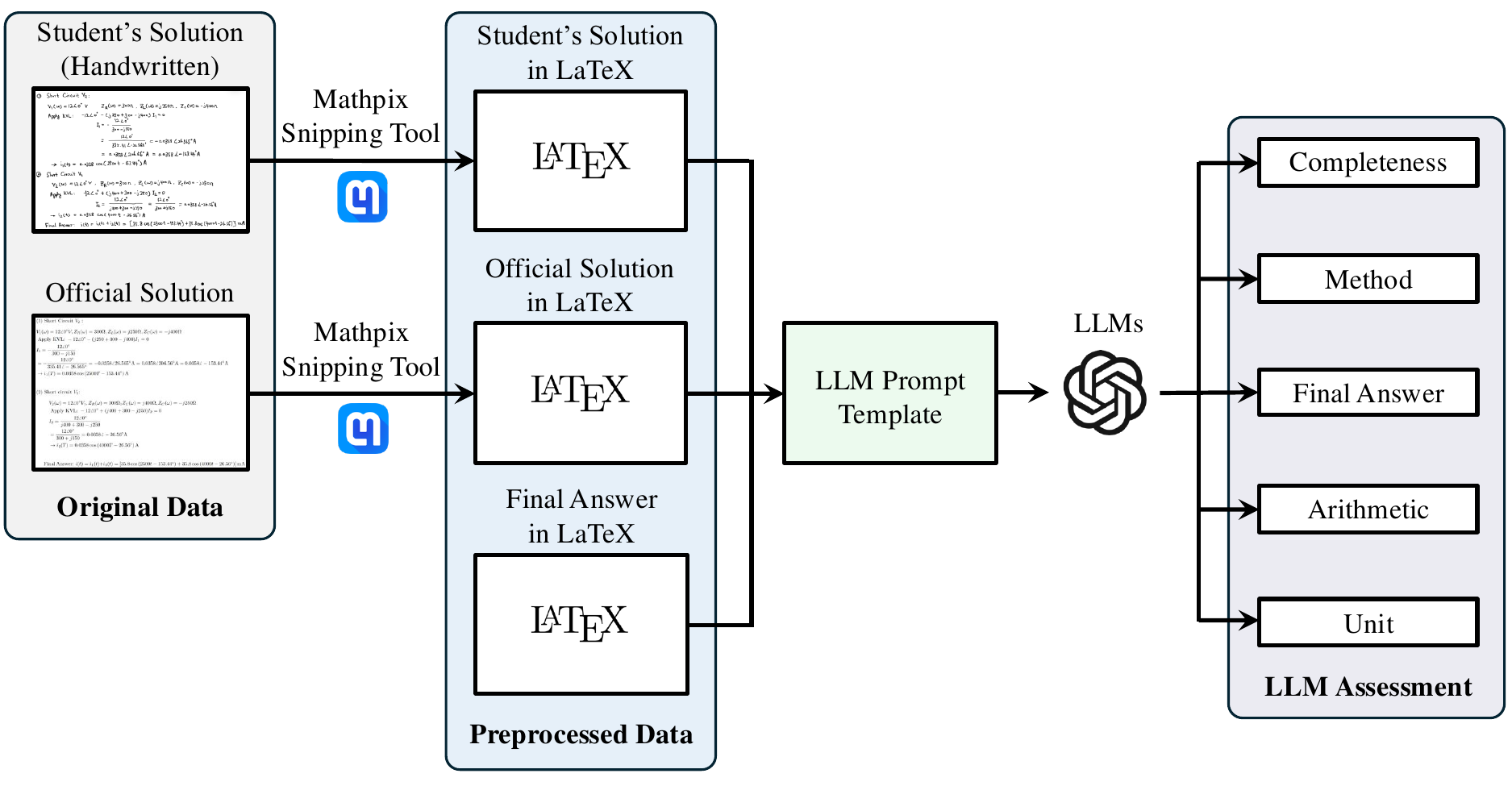}
	\caption{Methodological framework for the dataset collection and LLM evaluations}
	\label{fig:method}
\end{figure}

To the best of our knowledge, we are the first to select these metrics for evaluating LLM performance specifically within the domain of homework assessment for an undergraduate-level circuit analysis course. The prompt used in our assessment is closely tailored to these evaluation metrics. Fig. \ref{fig:method} shows the overall methodological framework for the dataset collection and LLM evaluations in this paper. 

\newpage
\section{LLM Evaluation Results}
\label{S4}
This section presents and analyzes the evaluation results of three LLMs on the dataset described in Section \ref{S31}. The tested LLMs include GPT-3.5 Turbo, GPT-4o, and Llama 3 70B\footnote{See the model descriptions at \url{https://platform.openai.com/docs/models/gpt-3-5-turbo} (GPT-3.5 Turbo), \url{https://platform.openai.com/docs/models/gpt-4o} (GPT-4o), and \url{https://github.com/meta-llama/llama3/blob/main/MODEL_CARD.md} (Llama 3 70B), respectively.}, which represent a basic baseline LLM, a state-of-the-art closed-source LLM, and a state-of-the-art open-source LLM, respectively. We leveraged the OpenAI API to get responses from GPT-3.5 Turbo and GPT-4o, and the Nvidia NIM API to obtain responses from Llama 3 70B\footnote{See \url{https://openai.com/api} (OpenAI API) and \url{https://build.nvidia.com/meta/llama3-70b} (Nvidia NIM API), respectively.}. To ensure the evaluation results are consistent and comparable, we used a common template for all three considered LLMs. Due to the free-form nature of the LLM responses, we \textit{manually} categorize the responses of LLMs into three classes: ``correct'', ``partially correct'', and ``incorrect'', based on the response's accuracy in assessing the student solution for each metric. The human rater evaluating the LLM responses is Liangliang Chen, the first author of this paper, who was also a teaching assistant for the circuit analysis course during the dataset collection period. In the assessment process, the rater carefully reviewed the LLM responses, categorizing them based on his expertise and the feedback he previously provided on students’ solutions in his role as a teaching assistant. Note that a response is considered partially correct if some sentences are accurate while others contain errors. Below, we present an example of a segment of a response classified as ``partially correct''. 

The response segment in the following box is used to evaluate whether the correct method is used in a student's solution that truly uses the correct method. The second sentence \ding{173} in this response is correct, while the third sentence \ding{174} is incorrect. In this case, we consider the method evaluation of the LLM as partially correct.

\begin{tcolorbox}[breakable, enhanced, width=\textwidth, opacityfill=1., colback=white]
	\ding{172} The student used some correct methods in their solution. \ding{173} They correctly calculated power $P$ using $P = VI$ and recognized that power is the rate of energy transfer. \ding{174} However, the method to calculate the final time $\Delta t$ was not clear or fully executed.
\end{tcolorbox}

Table \ref{Tab0} summarizes the percentages of correct responses from different LLMs across various topics. For the purposes of this summary, partially correct responses are treated the same as incorrect responses. The detailed classification results of partially correct and incorrect responses can be found in Tables \ref{Tab1}--\ref{Tab6}. The key observations are summarized as follows.

\begin{landscape}
	\begin{small}
		\begin{table}[!h]
			\centering
			\caption{Test results of different LLMs across various topics in circuit analysis}
				\begin{tabular}{ccccccccc}
					\hline
					\multirow{2}{*}{LLM} & \multirow{2}{*}{Topic} & \multirow{2}{*}{\# Data} & \multicolumn{6}{c}{Percentages of correct LLM responses in terms of different metrics} \\
					\cline{4-9}
					& & & Completeness & Method & Final Answer & Arithmetic & Unit & Average\\
					\hline
					\multirow{7}{*}{\makecell{GPT-3.5 Turbo\\\texttt{gpt-3.5-turbo-0125}}} & Var. \& Ele. & 40 & 55.00\% & 87.50\% & 47.50\% & 47.50\% & 72.50\% & 62.00\% \\ 
					& Resist. Cir. & 63 & 42.86\% & 73.02\% & 42.86\% & 23.81\% & 55.56\% & 47.62\%\\
					& Op. Amp. & 28 & 50.00\% & 78.57\% & 46.43\% & 28.57\% & 53.57\% & 51.43\%\\
					& Com. Resp. & 95 & 50.53\% & 69.47\% & 44.21\% & 35.79\% & 49.47\% & 49.89\%\\
					& Sinusoidal & 29 & 44.83\% & 86.21\% & 62.07\% & 58.62\% & 58.62\% & 62.07\%\\
					& Freq. Resp. & 28 & 42.66\% & 64.29\% & 53.57\% & 25.00\% & 21.43\% & 41.39\%\\\cline{2-9}
					& Average & --- & 48.04\% & 74.91\% & 47.35\% & 35.34\% & 52.65\% & 51.66\%\\
					\hline
					\multirow{7}{*}{\makecell{GPT-4o\\\texttt{gpt-4o-2024-05-13}}} & Var. \& Ele. & 40 & 95.00\% & \textbf{100.00\%} & \textbf{90.00\%} & \textbf{95.00\%} & \textbf{100.00\%} & \textbf{96.00\%}\\ 
					& Resist. Cir. & 63 & \textbf{95.24\%} & 92.06\% & 80.95\% & 80.95\% & \textbf{98.41\%} & 89.52\%\\
					& Op. Amp. & 28 & 78.57\% & \textbf{96.43\%} & 89.29\% & 82.14\% & \textbf{92.86\%} & 87.86\%\\
					& Com. Resp. & 95 & 89.47\% & 96.84\% & \textbf{80.00\%} & \textbf{80.00\%} & \textbf{86.32\%} & \textbf{86.53\%}\\
					& Sinusoidal & 29 & 68.97\% & 93.10\% & \textbf{82.76\%} & \textbf{75.86\%} & \textbf{68.97\%} & \textbf{77.93\%}\\
					& Freq. Resp. & 28 & \textbf{96.43\%} & 96.43\% & \textbf{92.86\%} & \textbf{92.86\%} & \textbf{92.86\%} & \textbf{94.29\%} \\\cline{2-9}
					& Average & --- & 89.05\% & 95.75\% & \textbf{84.10\%} & \textbf{83.39\%} & \textbf{90.46\%} & \textbf{88.55\%}\\
					\hline
					\multirow{7}{*}{\makecell{Llama 3 70B\\\texttt{llama3-70b-instruct}}} & Var. \& Ele. & 40 & \textbf{97.50\%} & \textbf{100.00\%} & 82.50\% & 75.00\% & 85.00\% & 88.00\%\\ 
					& Resist. Cir. & 63 & \textbf{95.24\%} & \textbf{98.41\%} & \textbf{90.48\%} & \textbf{82.54\%} & 92.00\% & \textbf{91.73\%}\\
					& Op. Amp. & 28 & \textbf{92.86\%} & 92.86\% & \textbf{92.86\%} & \textbf{89.29\%} & 89.29\% & \textbf{91.43\%}\\
					& Com. Resp. & 95 & \textbf{91.58\%} & \textbf{97.89\%} & 78.95\% & 73.68\% & 58.95\% & 80.21\%\\
					& Sinusoidal & 29 & \textbf{82.76\%} & \textbf{100.00\%} & \textbf{82.76\%} & 65.52\% & 55.17\% & 77.24\%\\
					& Freq. Resp. & 28 & \textbf{96.43\%} & \textbf{100.00\%} & 82.14\% & 89.29\% & 53.57\% & 84.29\%\\\cline{2-9}
					& Average & --- & \textbf{92.93\%} & \textbf{98.23\%} & \textbf{84.10\%} & 78.09\% & 72.07\% & 85.08\%\\
					\hline
					\multicolumn{9}{l}{\makecell[{{p{18.5cm}}}]{* \ding{172} Var. \& Ele. = Electric Circuit Variables and Elements; \ding{173} Resist. Cir. = Analysis of Resistive Circuits; \ding{174} Op. Amp. = The Operational Amplifier; \ding{175} Com. Resp. = The Complete Response of Circuits With Energy Storage Elements; \ding{176} Sinusoidal = Sinusoidal Steady-State Analysis; \ding{177} Freq. Resp. = Frequency Response.}}
				\end{tabular}
				\label{Tab0}
		\end{table}
	\end{small}
\end{landscape}

\begin{table}[htbp]
	\centering
	\label{Tab1}
	\caption{Test summary on the electric circuit variables and elements}
	\begin{tabular}{C{2.5cm}C{1.8cm}C{2.0cm}C{2.3cm}C{1.8cm}}
		\hline
		\multirow{2}{*}{LLM} & \multirow{2}{*}{Metric} & \multicolumn{3}{c}{Response Classification (Total Number of Data: 40)} \\
		\cline{3-5}
		& & Correct & Partially Correct & Incorrect \\
		\hline
		\multirow{5}{*}{\makecell{GPT-3.5 Turbo\\\texttt{gpt-3.5-turbo-0125}}} & Completeness & 22 (55.00\%) & 5 (12.50\%) & 13 (32.50\%) \\ 
		& Method & 35 (87.50\%) & 2 (5.00\%) & 3 (7.50\%) \\ 
		& Final Answer & 19 (47.50\%) & 14 (35.00\%) & 7 (17.50\%) \\
		& Arithmetic & 19 (47.50\%) & 5 (12.50\%) & 16 (40.00\%) \\
		& Unit & 29 (72.50\%) & 2 (5.00\%) & 9 (22.50\%) \\
		\hline
		\multirow{5}{*}{\makecell{GPT-4o\\\texttt{gpt-4o-2024-05-13}}} & Completeness & 38 (95.00\%) & 1 (2.50\%) & 1 (2.50\%) \\ 
		& Method & \textbf{40 (100.00\%)} & 0 (0.00\%) & 0 (0.00\%) \\ 
		& Final Answer & \textbf{36 (90.00\%)} & 4 (10.00\%) & 0 (0.00\%) \\
		& Arithmetic & \textbf{38 (95.00\%)} & 2 (5.00\%) & 0 (0.00\%) \\
		& Unit & \textbf{40 (100.00\%)} & 0 (0.00\%) & 0 (0.00\%) \\
		\hline
		\multirow{5}{*}{\makecell{Llama 3 70B\\\texttt{llama3-70b-instruct}}} & Completeness & \textbf{39 (97.50\%)} & 1 (2.50\%) & 0 (0.00\%) \\ 
		& Method & \textbf{40 (100.00\%)} & 0 (0.00\%) & 0 (0.00\%) \\ 
		& Final Answer & 33 (82.50\%) & 6 (15.00\%) & 1 (2.50\%) \\
		& Arithmetic & 30 (75.00\%) & 9 (22.50\%) & 1 (2.50\%) \\
		& Unit & 34 (85.00\%) & 4 (10.00\%) & 2 (5.00\%) \\
		\hline
		\multicolumn{5}{l}{\makecell[{{p{12cm}}}]{*The integers are the numbers of LLM responses that can categorized in the corresponding classes, while the percentages in parentheses represent the ratios of responses in the corresponding classes to the total number of data.}}
	\end{tabular}
\end{table}

\begin{table}[htbp]
	\centering
	\caption{Test summary on the analysis of resistive circuits}
	\begin{tabular}{C{2.5cm}C{1.8cm}C{2.0cm}C{2.3cm}C{1.8cm}}
		\hline
		\multirow{2}{*}{LLM} & \multirow{2}{*}{Metric} & \multicolumn{3}{c}{Response Classification (Total Number of Data: 63)} \\
		\cline{3-5}
		& & Correct & Partially Correct & Incorrect \\
		\hline
		\multirow{5}{*}{\makecell{GPT-3.5 Turbo\\\texttt{gpt-3.5-turbo-0125}}} & Completeness & 27 (42.86\%) & 12 (19.05\%) & 24 (38.10\%) \\ 
		& Method & 46 (73.02\%) & 5 (7.94\%) & 12 (19.05\%) \\ 
		& Final Answer & 27 (42.86\%) & 19 (30.16\%) & 17 (26.98\%) \\
		& Arithmetic & 15 (23.81\%) & 10 (15.87\%) & 38 (60.32\%) \\
		& Unit & 35 (55.56\%) & 7 (11.11\%) & 21 (33.33\%) \\
		\hline
		\multirow{5}{*}{\makecell{GPT-4o\\\texttt{gpt-4o-2024-05-13}}} & Completeness & \textbf{60 (95.24\%)} & 1 (1.59\%) & 2 (3.17\%) \\ 
		& Method & 58 (92.06\%) & 3 (4.76\%) & 2 (3.17\%) \\ 
		& Final Answer & 51 (80.95\%) & 8 (12.70\%) & 4 (6.35\%) \\
		& Arithmetic & 51 (80.95\%) & 9 (14.29\%) & 3 (4.76\%) \\
		& Unit & \textbf{62 (98.41\%)} & 1 (1.59\%) & 0 (0.00\%) \\
		\hline
		\multirow{5}{*}{\makecell{Llama 3 70B\\\texttt{llama3-70b-instruct}}} & Completeness & \textbf{60 (95.24\%)} & 1 (1.59\%) & 2 (3.17\%) \\ 
		& Method & \textbf{62 (98.41\%)} & 1 (1.59\%) & 0 (0.00\%) \\ 
		& Final Answer & \textbf{57 (90.48\%)} & 6 (9.52\%) & 0 (0.00\%) \\
		& Arithmetic & \textbf{52 (82.54\%)} & 11 (17.46\%) & 0 (0.00\%) \\
		& Unit & 58 (92.00\%) & 3 (4.76\%) & 2 (3.17\%) \\
		\hline
	\end{tabular}
	\label{Tab2}
\end{table}

\begin{table}[htbp]
	\centering
	\caption{Test summary on the operational amplifier}
	\begin{tabular}{C{2.5cm}C{1.8cm}C{2.0cm}C{2.3cm}C{1.8cm}}
		\hline
		\multirow{2}{*}{LLM} & \multirow{2}{*}{Metric} & \multicolumn{3}{c}{Response Classification (Total Number of Data: 28)} \\
		\cline{3-5}
		& & Correct & Partially Correct & Incorrect \\
		\hline
		\multirow{5}{*}{\makecell{GPT-3.5 Turbo\\\texttt{gpt-3.5-turbo-0125}}} & Completeness & 14 (50.00\%) & 2 (7.14\%) & 12 (42.86\%) \\ 
		& Method & 22 (78.57\%) & 1 (3.57\%) & 5 (17.86\%) \\ 
		& Final Answer & 13 (46.43\%) & 11 (39.29\%) & 4 (14.29\%) \\
		& Arithmetic & 8 (28.57\%) & 3 (10.71\%) & 17 (60.71\%) \\
		& Unit & 15 (53.57\%) & 7 (25.00\%) & 6 (21.43\%) \\
		\hline
		\multirow{5}{*}{\makecell{GPT-4o\\\texttt{gpt-4o-2024-05-13}}} & Completeness & 22 (78.57\%) & 5 (17.85\%) & 1 (3.57\%) \\ 
		& Method & 27 (96.43\%) & 1 (3.57\%) & 0 (0.00\%) \\ 
		& Final Answer & 25 (89.29\%) & 1 (3.57\%) & 2 (7.14\%) \\
		& Arithmetic & 23 (82.14\%) & 3 (10.71\%) & 2 (7.14\%) \\
		& Unit & \textbf{26 (92.86\%)} & 1 (3.57\%) & 1 (3.57\%) \\
		\hline
		\multirow{5}{*}{\makecell{Llama 3 70B\\\texttt{llama3-70b-instruct}}} & Completeness & \textbf{26 (92.86\%)} & 1 (3.57\%) & 1 (3.57\%) \\ 
		& Method & \textbf{26 (92.86\%)} & 1 (3.57\%) & 1 (3.57\%) \\ 
		& Final Answer & \textbf{26 (92.86\%)} & 2 (7.14\%) & 0 (0.00\%) \\
		& Arithmetic & \textbf{25 (89.29\%)} & 1 (3.57\%) & 2 (7.14\%) \\
		& Unit & 25 (89.29\%) & 3 (10.71\%) & 0 (0.00\%) \\
		\hline
	\end{tabular}
	\label{Tab3}
\end{table}

\begin{table}[htbp]
	\centering
	\caption{Test summary on the complete response of circuits with energy storage elements}
	\begin{tabular}{C{2.5cm}C{1.8cm}C{2.0cm}C{2.3cm}C{1.8cm}}
		\hline
		\multirow{2}{*}{LLM} & \multirow{2}{*}{Metric} & \multicolumn{3}{c}{Response Classification (Total Number of Data: 95)} \\
		\cline{3-5}
		& & Correct & Partially Correct & Incorrect \\
		\hline
		\multirow{5}{*}{\makecell{GPT-3.5 Turbo\\\texttt{gpt-3.5-turbo-0125}}} & Completeness & 48 (50.53\%) & 0 (0.00\%) & 47 (49.47\%) \\ 
		& Method & 66 (69.47\%) & 9 (9.47\%) & 20 (21.05\%) \\ 
		& Final Answer & 42 (44.21\%) & 28 (29.47\%) & 25 (26.32\%) \\
		& Arithmetic & 34 (35.79\%) & 14 (14.74\%) & 47 (49.47\%) \\
		& Unit & 47 (49.47\%) & 26 (27.37\%) & 22 (23.16\%) \\
		\hline
		\multirow{5}{*}{\makecell{GPT-4o\\\texttt{gpt-4o-2024-05-13}}} & Completeness & 85 (89.47\%) & 7 (7.37\%) & 3 (3.16\%) \\ 
		& Method & 92 (96.84\%) & 3 (3.16\%) & 0 (0.00\%) \\ 
		& Final Answer & \textbf{76 (80.00\%)} & 12 (12.63\%) & 7 (7.37\%) \\
		& Arithmetic & \textbf{76 (80.00\%)} & 11 (11.58\%) & 8 (8.42\%) \\
		& Unit & \textbf{82 (86.32\%)} & 10 (10.53\%) & 3 (3.16\%) \\
		\hline
		\multirow{5}{*}{\makecell{Llama 3 70B\\\texttt{llama3-70b-instruct}}} & Completeness & \textbf{87 (91.58\%)} & 3 (3.16\%) & 5 (5.26\%) \\ 
		& Method & \textbf{93 (97.89\%)} & 2 (2.11\%) & 0 (0.00\%) \\ 
		& Final Answer & 75 (78.95\%) & 16 (16.84\%) & 4 (4.21\%) \\
		& Arithmetic & 70 (73.68\%) & 16 (16.84\%) & 9 (9.47\%) \\
		& Unit & 56 (58.95\%) & 30 (31.58\%) & 9 (9.47\%) \\
		\hline
	\end{tabular}
	\label{Tab4}
\end{table}

\begin{table}[htbp]
	\centering
	\caption{Test summary on the sinusoidal steady-state analysis}
	\begin{tabular}{C{2.5cm}C{1.8cm}C{2.0cm}C{2.3cm}C{1.8cm}}
		\hline
		\multirow{2}{*}{LLM} & \multirow{2}{*}{Metric} & \multicolumn{3}{c}{Response Classification (Total Number of Data: 29)} \\
		\cline{3-5}
		& & Correct & Partially Correct & Incorrect \\
		\hline
		\multirow{5}{*}{\makecell{GPT-3.5 Turbo\\\texttt{gpt-3.5-turbo-0125}}} & Completeness & 13 (44.83\%) & 1 (3.45\%) & 15 (51.72\%) \\ 
		& Method & 25 (86.21\%) & 0 (0.00\%) & 4 (13.79\%) \\ 
		& Final Answer & 18 (62.07\%) & 8 (27.59\%) & 3 (10.34\%) \\
		& Arithmetic & 17 (58.62\%) & 3 (10.34\%) & 9 (31.03\%) \\
		& Unit & 17 (58.62\%) & 3 (10.34\%) & 9 (31.03\%) \\
		\hline
		\multirow{5}{*}{\makecell{GPT-4o\\\texttt{gpt-4o-2024-05-13}}} & Completeness & 20 (68.97\%) & 6 (20.69\%) & 3 (10.34\%) \\ 
		& Method & 27 (93.10\%) & 1 (3.45\%) & 1 (3.45\%) \\ 
		& Final Answer & \textbf{24 (82.76\%)} & 2 (6.90\%) & 3 (10.34\%) \\
		& Arithmetic & \textbf{22 (75.86\%)} & 4 (13.79\%) & 3 (10.34\%) \\
		& Unit & \textbf{20 (68.97\%)} & 4 (13.79\%) & 5 (17.24\%) \\
		\hline
		\multirow{5}{*}{\makecell{Llama 3 70B\\\texttt{llama3-70b-instruct}}} & Completeness & \textbf{24 (82.76\%)} & 3 (10.34\%) & 2 (6.90\%) \\ 
		& Method & \textbf{29 (100.00\%)} & 0 (0.00\%) & 0 (0.00\%) \\ 
		& Final Answer & \textbf{24 (82.76\%)} & 3 (10.34\%) & 2 (6.90\%) \\
		& Arithmetic & 19 (65.52\%) & 5 (17.24\%) & 5 (17.24\%) \\
		& Unit & 16 (55.17\%) & 6 (20.69\%) & 7 (24.14\%) \\
		\hline
	\end{tabular}
	\label{Tab5}
\end{table}

\begin{table}[htbp]
	\centering
	\caption{Test summary on the frequency response}
	\begin{tabular}{C{2.5cm}C{1.8cm}C{2.0cm}C{2.3cm}C{1.8cm}}
		\hline
		\multirow{2}{*}{LLM} & \multirow{2}{*}{Metric} & \multicolumn{3}{c}{Response Classification (Total Number of Data: 28)} \\
		\cline{3-5}
		& & Correct & Partially Correct & Incorrect \\
		\hline
		\multirow{5}{*}{\makecell{GPT-3.5 Turbo\\\texttt{gpt-3.5-turbo-0125}}} & Completeness & 12 (42.66\%) & 3 (10.71\%) & 13 (46.43\%) \\ 
		& Method & 18 (64.29\%) & 0 (0.00\%) & 10 (35.71\%) \\ 
		& Final Answer & 15 (53.57\%) & 2 (7.14\%) & 11 (39.29\%) \\
		& Arithmetic & 7 (25.00\%) & 4 (14.29\%) & 17 (60.71\%) \\
		& Unit & 6 (21.43\%) & 19 (67.86\%) & 3 (10.71\%) \\
		\hline
		\multirow{5}{*}{\makecell{GPT-4o\\\texttt{gpt-4o-2024-05-13}}} & Completeness & \textbf{27 (96.43\%)} & 0 (0.00\%) & 1 (3.57\%) \\ 
		& Method & 27 (96.43\%) & 1 (3.57\%) & 0 (0.00\%) \\ 
		& Final Answer & \textbf{26 (92.86\%)} & 1 (3.57\%) & 1 (3.57\%) \\
		& Arithmetic & \textbf{26 (92.86\%)} & 1 (3.57\%) & 1 (3.57\%) \\
		& Unit & \textbf{26 (92.86\%)} & 2 (7.14\%) & 0 (0.00\%) \\
		\hline
		\multirow{5}{*}{\makecell{Llama 3 70B\\\texttt{llama3-70b-instruct}}} & Completeness & \textbf{27 (96.43\%)} & 0 (0.00\%) & 1 (3.57\%) \\ 
		& Method & \textbf{28 (100.00\%)} & 0 (0.00\%) & 0 (0.00\%) \\ 
		& Final Answer & 23 (82.14\%) & 1 (3.57\%) & 4 (14.29\%) \\
		& Arithmetic & 25 (89.29\%) & 1 (3.57\%) & 2 (7.14\%) \\
		& Unit & 15 (53.57\%) & 7 (25.00\%) & 6 (21.43\%) \\
		\hline
	\end{tabular}
	\label{Tab6}
\end{table}

\begin{itemize}
\item [i)] GPT-4o and Llama 3 70B significantly outperform GPT-3.5 Turbo across all metrics. Notably, GPT-4o achieves correct response rates exceeding 80\% on average across all metrics. This highlights the advancements in newly developed LLMs for homework assessment in the circuit analysis course. 

\item [ii)] GPT-4o demonstrates more consistent performance than Llama 3 70B, with a minimum correct response rate of 68.97\% across all topics and metrics. In contrast, Llama 3 70B falls below 60\% on the unit metric for three relatively advanced topics. 

\item [iii)] In the metrics of completeness and method, Llama 3 70B appears to outperform GPT-4o. Detailed analysis shows that Llama 3 70B tends to give positive evaluations of students' solutions, whereas GPT-4o is more rigorous. Consequently, GPT-4o has a higher false negative rate, while Llama 3 70B has a higher false positive rate. Since most students complete their solutions and use correct methods, GPT-4o is more likely to provide incorrect or partially incorrect responses for these two metrics. However, because students are more prone to omitting units, Llama 3 70B performs significantly worse than GPT-4o in the unit metric.

\item [iv)] The performance of LLMs varies across topics with different levels of difficulty. Sinusoidal steady-state analysis is the most challenging topic for both GPT-4o and Llama 3 70B, as indicated by the average percentage of correct responses. We speculate that this difficulty arises because steady-state analysis involves calculations and conversions of values across different forms, including complex numbers, phasors, and sinusoids. 
\end{itemize}

Although GPT-4o shows the most promising evaluation results according to Table \ref{Tab0} and the analyses above, it is important to summarize the errors it frequently makes. Table \ref{TabGPT4o} lists the types of errors GPT-4o tends to make along with associated examples. Appendix \ref{SD2} presents additional notable example responses from GPT-4o and Llama 3 70B across various topics, aiming to better illustrate the strengths and weaknesses of these different LLMs. Future research will focus on mitigating these issues. 

\begin{center}
	\begin{small}
	\begin{longtable}{ccc}
		\caption{Notable examples of GPT-4o's response} \label{TabGPT4o} \\
		\hline
		Type & Example & Note \\
		\hline
		\multirow{6}{*}{\makecell{Unit\\Conversion}} & \makecell[{{p{6.0cm}}}]{The correct cost should be \$0.264, but your final answer is 26.4 cents (i.e., \$0.264). It seems you made an error when interpreting units (hours should have been used instead of seconds).} & \makecell[{{p{4.0cm}}}]{GPT-4o fails to identify the equivalence between 26.4 cents and \$0.264.} \\ \cline{2-3}
		& \makecell[{{p{6.0cm}}}]{The units for power are inconsistent and incorrect due to the sign convention error. Additionally, the answer is expected in milliwatts (mW), but the unit was given in watts (W) which adds to the confusion.} & \makecell[{{p{4.0cm}}}]{GPT-4o is overly confident that the answer is expected in milliwatts. In fact, the answer given in watts is also acceptable.} \\ \hline
		\multirow{10}{*}{\makecell{False\\Incomplete-\\ness}} & \makecell[{{p{6.0cm}}}]{The student's solution is incomplete. While they calculated the voltage across the resistors and the power absorbed by each resistor, they did not address the direction of current and adherence to the passive sign convention.} & \makecell[{{p{4.0cm}}}]{The student's solution should be considered complete since they calculated the voltage across the resistors and the power absorbed by each resistor.} \\\cline{2-3}
		& \makecell[{{p{6.0cm}}}]{The student's solution is incomplete. While it attempts to find the current ($i$) and voltage ($v_o$), it lacks a detailed step-by-step explanation which includes the intermediate steps and proper application of Kirchhoff's Voltage Law (KVL) and Kirchhoff's Current Law (KCL). The discussion concerning $i_2$ and intermediate voltages like $v_a$ is missing.} & \makecell[{{p{4.0cm}}}]{The student's solution is actually complete but misses some step-by-step explanations.} \\
		\hline
		\multirow{7}{*}{\makecell{Number\\Format}} & \makecell[{{p{6.0cm}}}]{Unfortunately, your final answer for \( R_t \) is incorrect. The official answer is \( R_t = -\frac{8}{3} \, \Omega \), but you calculated it as \( -\frac{16}{6} \, \Omega \).} & \makecell[{{p{4.0cm}}}]{GPT-4o fails to recognize the equivalence between $-\frac{8}{3}~\Omega$ and $-\frac{16}{6}~\Omega$.} \\
		\cline{2-3}
		& \makecell[{{p{6.0cm}}}]{The student's final answer is: $V(t)=\sqrt{74} \cos \left(8 t-9.462^{\circ}\right)$\\
			While the correct answer should be: $v(t)=8.602 \cos \left(8 t-9.462^{\circ}\right) ~\text{V}$\\
			It can be observed that the student's amplitude calculation is incorrect.} & \makecell[{{p{4.0cm}}}]{Since \(\sqrt{74} \approx 8.602\), the student's final answer matches the official solution up to a rounding error.} \\
		\hline
		\makecell{Alternative\\Method} & \makecell[{{p{6.0cm}}}]{The student applies KCL to find \(R_t\). However, the method and steps are somewhat unclear and differ from the standard approach usually taught, which directly utilizes the short-circuit current \(i_{sc}\) and open-circuit voltage \(v_{oc}\). This might be confusing.} & \makecell[{{p{4.0cm}}}]{The student's solution uses an alternative approach to find $R_{t}$. Although the method used is different from the approach in the official solution, it is also correct.} \\
		\hline
		\makecell{Different\\Notations} & \makecell[{{p{6.0cm}}}]{There are arithmetic errors in your calculations for \( i_1 \) and \( i_2 \). \\
				-- Correct calculation:\\$i_1=\frac{8}{16+8}(-6)=-2 \text{ A}$, $i_2=\frac{8}{8+8}(-6)=-3 \text{ A}$\\
				-- Your incorrect calculation:\\$i_1=\frac{1 / 8}{\frac{1}{8}+\frac{1}{8}} \cdot 6 \text{ A}=3 \text{ A}$, $i_2=\frac{1 / 16}{ \frac{1}{16}+\frac{1}{8}} \cdot 6 \text{ A}=2 \text{ A}
				$} & \makecell[{{p{4.0cm}}}]{GPT-4o fails to recognize that the student's solution uses a different notation system from the official solution. The currents \( i_{1} \) and \( i_{2} \) in the student's solution represent \( i_{2} \) and \( i_{1} \), respectively, in the official solution. GPT-4o cannot recognize the exchanged variable notations in this example.} \\
		\hline
		\multirow{29}{*}{\makecell{Arithmetic}} & \makecell[{{p{6.0cm}}}]{There are errors in the arithmetic and the process leading to the calculation, primarily in their steps involving the arithmetic for current division:\\
				-- For the 9 A source: $i_{20A} = 9\text{~A} \cdot \frac{1/35\Omega}{1/10\Omega + 1/35\Omega} \neq 2\text{~A}$\\
				Correct method: $i_{20A} = 9\text{~A} \cdot \frac{10\Omega}{10\Omega + 35\Omega} = 9\text{~A} \cdot \frac{10}{45} = 9\text{~A} \cdot \frac{2}{9} = 2\text{~A}$\\
				-- For the 6 A source: $i_{20B} = 6\text{~A} \cdot \frac{1/30\Omega}{1/30\Omega + 1/15\Omega} \neq 2\text{~A}$\\
				Correct method: $i_{20B} = 6\text{~A} \cdot \frac{15\Omega}{15\Omega + 30\Omega} = 6\text{~A} \cdot \frac{15}{45} = 6\text{~A} \cdot \frac{1}{3} = 2\text{~A}$} & \makecell[{{p{4.0cm}}}]{Although GPT-4o sometimes performs detailed mathematical calculations, the results are not necessarily correct. In this example, we should have $9\text{~A} \cdot \frac{1/35\Omega}{1/10\Omega + 1/35\Omega} = 2\text{~A}$ and $6\text{~A} \cdot \frac{1/30\Omega}{1/30\Omega + 1/15\Omega} = 2\text{~A}$.} \\
		\cline{2-3}
		& \makecell[{{p{6.0cm}}}]{There are no significant arithmetic errors that affect the final result. However, there is a sign issue to note:\\
				-- In the equation for \( \text{KCL@} V_b \), you have correctly identified the expression \( \frac{-V_3}{6000}-\frac{V_0}{6000}=0 \), but then you incorrectly simplified it to \( -4 - V_0 = 0 \). The correct simplification should be \( -\frac{4}{6000} - \frac{V_0}{6000} = 0 \) which equates to \( -4 - V_0 = 0 \), so this appears to be more of a transcription error than an arithmetic error.} & \makecell[{{p{4.0cm}}}]{The response of GPT-4o is redundant since it is completely fine to derive \( -4 - V_0 = 0 \) directly from \( \frac{-V_3}{6000} - \frac{V_0}{6000} = 0 \), given \( V_{3} = 4 \) V.} \\
		\cline{2-3}
		& \makecell[{{p{6.0cm}}}]{The student's final expression for \( V_0 \) is: $V_0 = \frac{R_1(V_1 - V_2) - R_3(V_2 - V_1)}{R_2} + V_1 - V_2$\\
				Simplifying this, \\
				$V_0 = \frac{R_1(V_1 - V_2) + R_3(V_1 - V_2)}{R_2} + V_1 - V_2$, 
				$V_0 = \frac{(R_1 + R_3)(V_1 - V_2)}{R_2} + V_1 - V_2$\\
				This does not match the official solution: $v_0 = \frac{R_1 + R_2 + R_3}{R_2}(v_1 - v_2)$\\
				Thus, the student's final answer is incorrect. The steps might have errors that cause the final discrepancy.} & \makecell[{{p{4.0cm}}}]{GPT-4o has some level of calculation ability. However, it does not fully simplify the original equation for \( V_{0} \) and thus made an incorrect conclusion that the student's final expression does not match the official solution.} \\
		\cline{2-3}
		& \makecell[{{p{6.0cm}}}]{The student's final answer $i(t) = -3 \sin(2t + 30^\circ)$ is not in the correct form. The official solution is given by $i(t) = 3 \cos(2t + 120^\circ)$. There is an error in converting the trigonometric function from sine to cosine form.} & \makecell[{{p{4.0cm}}}]{GPT-4o does not identify the equivalence between \(-3 \sin(2t + 30^\circ)\) and \(3 \cos(2t + 120^\circ)\).} \\
		\hline
		\makecell{Rounding\\Errors} & \makecell[{{p{6.0cm}}}]{The student's final answer \( 14.69 \angle 5.55^{\circ} \) is close but not correct. The correct answer is \( 14.67 \angle 5.6^{\circ} \).} & \makecell[{{p{4.0cm}}}]{GPT-4o does not recognize the value discrepancies in the student's final answer as rounding errors.} \\
		\hline
		\makecell{Term\\Order in\\Equations} & \makecell[{{p{6.0cm}}}]{The final equation \( 1 \times 10^8 v_s = \frac{d^2 v}{dt^2} + 3000 \frac{dv}{dt} + 1.02 \times 10^8 v_a \) the student reaches is incorrect primarily because of coefficient and term sign errors throughout the derivation. This divergence from the correct answer \( 1 \times 10^8 v_{\mathrm{s}} = 1.02 \times 10^8 v + 3000 \frac{\mathrm{d}v}{\mathrm{d}t} + \frac{\mathrm{d}^2 v}{\mathrm{d}t^2} \) indicates a need to revisit algebraic and differential handling within this setup.} & \makecell[{{p{4.0cm}}}]{GPT-4o fails to recognize the equivalence between two differential equations in which the terms are arranged in different orders.} \\
		\hline
		\makecell{Hallucina-\\tions} & \makecell[{{p{6.0cm}}}]{-- Consistently use proper notation for complex numbers and phasors.\\
				-- Provide final results in correct units — Volts for $\mathbf{V}_{\mathrm{oc}}$ and Ohms for $\mathbf{Z}_{\mathrm{t}}$. } & \makecell[{{p{4.0cm}}}]{The student, in effect, did not consistently use units for all variables.} \\
		\hline
	\end{longtable}
	\end{small}
\end{center} 

\section{Limitations of LLMs in Circuit Analysis}
\label{S5}
Apart from the \textit{homework assessment} task discussed in Section \ref{S4}, this section discusses the limitations of current state-of-the-art LLMs/VLMs in \textit{solving circuit analysis problems}, using the results of GPT-4o as examples. On the one hand, the uncovered limitations can inform the future training of general LLMs or the fine-tuning of specific language models. On the other hand, they provide guidance on how to leverage the advantages and avoid the disadvantages of the currently available LLMs, which also shape the experiment design in this paper.

\subsection{Circuit Diagram Recognition and Understanding}
Circuit diagrams are key components in circuit analysis and the associated homework and solutions. Ideally, if a VLM can recognize the elements in a circuit diagram and understand their connections correctly, we can leverage this capability to extract the relevant prior knowledge from the VLM to provide useful instructions to students.

Unfortunately, GPT-4o, one of the state-of-the-art multimodal language models, cannot interpret a circuit diagram reliably. The following example shows that GPT-4o provides unreliable information when asked to list the elements in a circuit diagram and explain how these elements are connected. For the responses of GPT-4o in this section, we mark the incorrect parts with a \colorbox{red}{red} background and the dubious contents with a \colorbox{yellow}{yellow} background. We additionally mark with the \colorbox{green}{green} background the response segments that we would like to highlight as correct.

\vspace{0.25cm}
\noindent
\img{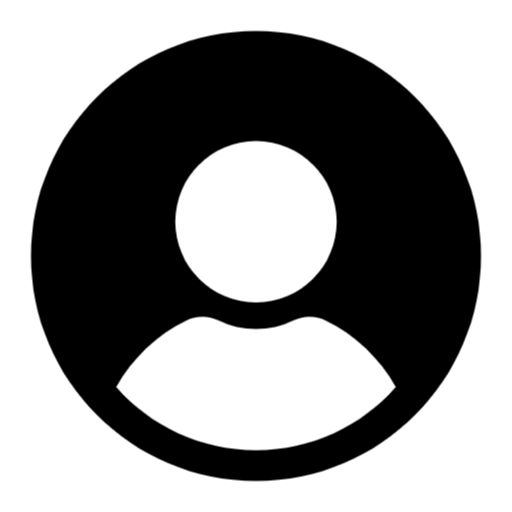} User

\begin{center}
	\begin{tcolorbox}[breakable, enhanced, width=\textwidth, opacityfill=1., colback=white]
		\begin{center}
			\frame{\includegraphics[width=0.5\textwidth]{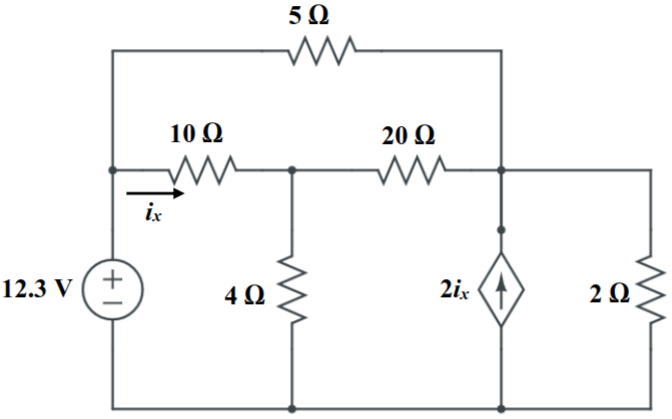}}
		\end{center}
		
		Now you play the role of an expert in circuit analysis. Can you explain how many elements are in the attached circuit diagram and how these elements are connected with each other? 
	\end{tcolorbox}
\end{center}

\hfill GPT-4o \img{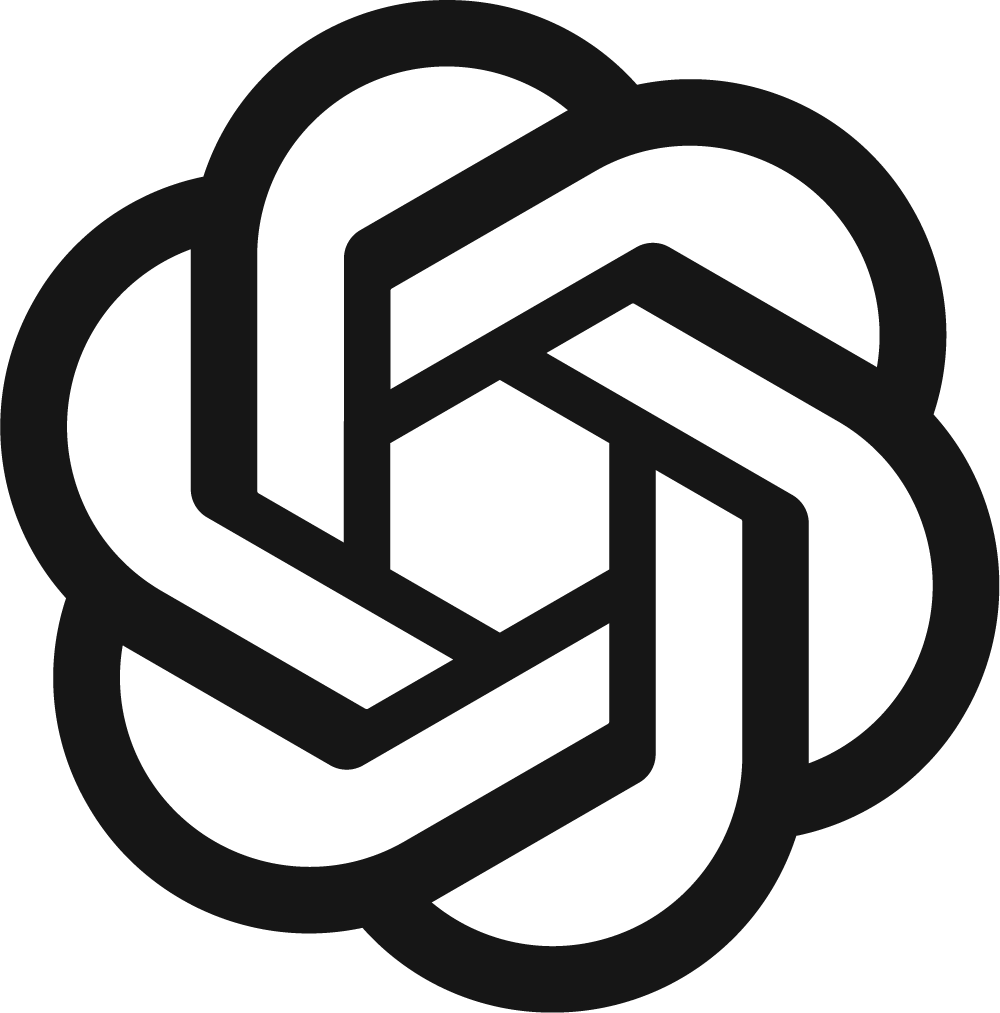}
\begin{center}
	\begin{tcolorbox}[breakable, enhanced, width=\textwidth, opacityfill=1., colback=white]
		The given circuit diagram consists of several elements, and they are connected as follows:\\
		
		\textbf{Elements:}\\
		\begin{itemize}
			\item [1.] Voltage Source:
			\begin{itemize}
				\item One voltage source of 12.3 V.
			\end{itemize}
			
			\item [2.] Resistors:
			\begin{itemize}
				\item \colorbox{red}{Six resistors} with the following values:
				\item 5 $\Omega$
				\item 10 $\Omega$
				\item 20 $\Omega$
				\item 4 $\Omega$
				\item 2 $\Omega$ (\colorbox{red}{two resistors} of this value)
			\end{itemize}
			
			\item [3.] Dependent Current Source:
			\begin{itemize}
				\item One dependent current source with a value of \(2i_x\), where \(i_x\) is the current through the 10 $\Omega$ resistor.
			\end{itemize}
		\end{itemize}
		~
		
		\textbf{Connections:}\\
		\begin{itemize}
			\item [1.] Voltage Source:
			\begin{itemize}
				\noindent\hspace*{-8pt}
				\colorbox{yellow}{
					\begin{minipage}{0.841\textwidth} 
						\item The positive terminal of the 12.3 V voltage source is connected to one end of the 4 $\Omega$ resistor and the 10 $\Omega$ resistor.
				\end{minipage}}
			\end{itemize}
			
			\item [2.] 10 $\Omega$ Resistor:
			\begin{itemize}
				\item One end is connected to the positive terminal of the voltage source, and the other end is connected to the 20 $\Omega$ resistor, \colorbox{red}{where the current \(i_x\) is defined}.
			\end{itemize}
			
			\item [3.] 20 $\Omega$ Resistor:
			\begin{itemize}
				\item One end is connected to the 10 $\Omega$ resistor, and the other end is connected to a node that connects to the dependent current source and one of the 2 $\Omega$ resistors.
			\end{itemize}
			
			\item [4.] 4 $\Omega$ Resistor:
			\begin{itemize}
				\item \colorbox{red}{One end is connected to the positive terminal of the voltage source,} and the other end is connected to the bottom node (ground).
			\end{itemize}
			
			\item [5.] 5 $\Omega$ Resistor:
			\begin{itemize}
				\noindent\hspace*{-8pt}
				\colorbox{red}{
					\begin{minipage}{0.841\textwidth}
						\item Connected between the top node (the junction of the 10 $\Omega$ and 20 $\Omega$ resistors) and the bottom node (ground).
				\end{minipage}}
			\end{itemize}
			
			\item [6.] Dependent Current Source (\(2i_x\)):
			\begin{itemize}
				\item The \colorbox{red}{positive terminal} is connected to the bottom node (ground), and the negative terminal is connected to a node that also connects to the 20 $\Omega$ resistor and one of the 2 $\Omega$ resistors.
			\end{itemize}
			
			\item [7.] First 2 $\Omega$ Resistor:
			\begin{itemize}
				\item One end is connected to the node where the 20 $\Omega$ resistor and the dependent current source meet, and the other end is connected to the bottom node (ground).
			\end{itemize}
			
			\item [8.] \colorbox{red}{Second 2 $\Omega$ Resistor:}
			\begin{itemize}
				\noindent\hspace*{-8pt}
				\colorbox{red}{
					\begin{minipage}{0.841\textwidth}
						\item One end is connected to the node where the 4 $\Omega$ resistor connects to the voltage source and the other end is connected to the bottom node (ground).
				\end{minipage}}
			\end{itemize}
		\end{itemize}
		The circuit consists of \colorbox{red}{six resistors}, one independent voltage source, and one dependent current source. The resistors and sources are interconnected as described, forming a network with multiple branches and nodes.
	\end{tcolorbox}
\end{center}

In this example, GPT-4o is asked to identify the elements in an attached circuit diagram and explain how the elements are connected. In its response, GPT-4o mistakenly lists a non-existing resistor due to hallucinations. Regarding the element connections, GPT-4o's explanations are dubious and often incorrect. Since these diagrams contain most of the problem information\footnote{In this example, the text portion of the problem description is simply, ``Determine $i_{x}$ in the circuit below.''}, GPT-4o lacks the necessary context to evaluate a student's solution accurately if no additional information about the specific problem is provided. This is where the reference solutions, along with their final answers, play a critical role in the LLM prompt in Section \ref{S32}. They offer detailed information for specific problems and serve as reference benchmarks for homework assessment, enabling LLMs to understand what each problem requires without having to infer it. The failure of GPT-4o to recognize circuit diagrams also implies that we cannot use it to directly solve undergraduate circuit analysis problems, considering that most problem statements include one or more circuit diagrams. We further tested this aspect in the following example.

\vspace{0.2cm}
\noindent
\img{figures/user_logo.png} User

\begin{center}
	\begin{tcolorbox}[breakable, enhanced, width=\textwidth, opacityfill=1., colback=white]
		\begin{center}
			\frame{\includegraphics[width=0.5\textwidth]{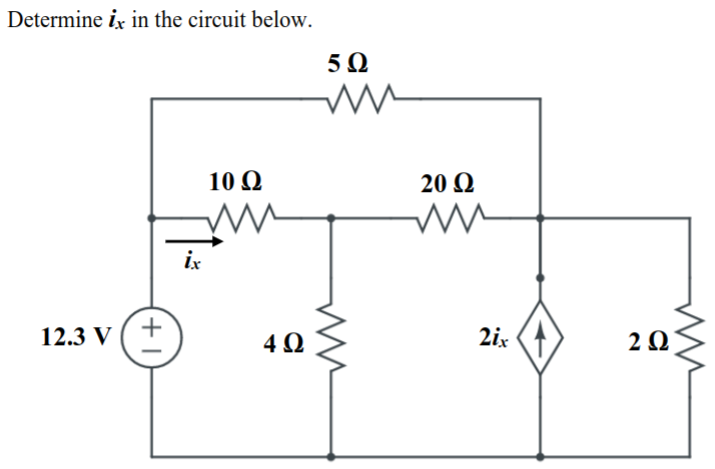}}
		\end{center}
		Now you play the role of an expert in circuit analysis. Can you solve the circuit analysis problem in the attached image? 
	\end{tcolorbox}
\end{center}

\hfill GPT-4o \img{figures/gpt_logo.png}
\begin{center}
	\begin{tcolorbox}[breakable, enhanced, width=\textwidth, opacityfill=1., colback=white]
		Sure, I can help you analyze this circuit. Let's follow these steps to find \( i_x \):\\
		
		\begin{itemize}
			\item [1.] Label the Nodes and Write Node-Voltage Equations: 
			\begin{itemize}
				\item Let's label the node where the 12.3V source is connected as \( V_1 \).
				\item Let's label the node where the 10 $\Omega$ and 20 $\Omega$ resistors meet as \( V_2 \).
				\item Let the voltage at the bottom of the circuit be 0V (ground).
			\end{itemize}
			
			\item [2.] Apply Kirchhoff's Current Law (KCL) at each Node: 
			\begin{itemize}
				\item At node \( V_1 \): 
				
				\noindent\hspace*{38pt}
				\colorbox{red}{
					\begin{minipage}{0.6\textwidth}
						\[
						\frac{V_1 - 12.3}{10} + \frac{V_1 - V_2}{4} + \frac{V_1}{5} = 0
						\]
				\end{minipage}}
				
				\item At node \( V_2 \): 
				
				\noindent\hspace*{38pt}
				\colorbox{red}{
					\begin{minipage}{0.6\textwidth}
						\[
						\frac{V_2 - V_1}{4} + \frac{V_2}{20} + \frac{V_2}{2} = 2i_x
						\]
				\end{minipage}}
				~
				
				Since \colorbox{red}{\(i_x = \frac{V_1 - 12.3}{10}\)}, substitute \( i_x \) in the above equation.\\
				
				\noindent\hspace*{38pt}
				\colorbox{yellow}{
					\begin{minipage}{0.6\textwidth}
						\[
						\colorbox{yellow}{$2i_x = 2 \cdot \frac{V_1 - 12.3}{10}$}
						\]
				\end{minipage}}
			\end{itemize}
			
			\item [3.] \colorbox{yellow}{
				Simplify and Solve the Equations:}
			
			\begin{itemize}
				\item \colorbox{yellow}{First Equation:}
				
				\noindent\hspace*{38pt}
				\colorbox{yellow}{
					\begin{minipage}{0.6\textwidth}
						\[
						\frac{V_1 - 12.3}{10} + \frac{V_1 - V_2}{4} + \frac{V_1}{5} = 0
						\]
				\end{minipage}}
				~
				
				\colorbox{yellow}{Simplifying,}
				
				\noindent\hspace*{38pt}
				\colorbox{yellow}{
				\begin{minipage}{0.6\textwidth}
					\[
					\frac{V_1 - 12.3}{10} + \frac{V_1 - V_2}{4} + \frac{V_1}{5} = 0
					\]
					
					\[
					\frac{V_1 - 12.3}{10} + \frac{V_1 - V_2}{4} + \frac{2V_1}{10} = 0
					\]
				\end{minipage}}
			
				\noindent\hspace*{38pt}
				\colorbox{red}{
					\begin{minipage}{0.6\textwidth}
						\[
						\frac{V_1 - 12.3}{10} + \frac{2.5V_1 - 2.5V_2}{10} = 0
						\]
				\end{minipage}}
			
				\noindent\hspace*{38pt}
				\colorbox{yellow}{
				\begin{minipage}{0.6\textwidth}
					\[
					V_1 - 12.3 + 2.5V_1 - 2.5V_2 = 0
					\]
					
					\[
					3.5V_1 - 2.5V_2 = 12.3
					\]
					
					\[
					7V_1 - 5V_2 = 24.6 \quad \text{(Equation 1)}
					\]
			\end{minipage}}
				
				\item \colorbox{yellow}{Second Equation:} 
				
				\noindent\hspace*{38pt}
				\colorbox{yellow}{
					\begin{minipage}{0.6\textwidth}
						\[
						\frac{V_2 - V_1}{4} + \frac{V_2}{20} + \frac{V_2}{2} = 2 \cdot \frac{V_1 - 12.3}{10}
						\]
				\end{minipage}}
				
				\colorbox{yellow}{Simplifying,}
				
				\noindent\hspace*{38pt}
				\colorbox{yellow}{
					\begin{minipage}{0.6\textwidth}
						\[
						\frac{V_2 - V_1}{4} + \frac{V_2}{20} + \frac{V_2}{2} = \frac{2(V_1 - 12.3)}{10}
						\]
						
						\[
						\frac{V_2 - V_1}{4} + \frac{V_2}{20} + \frac{V_2}{2} = \frac{V_1 - 12.3}{5}
						\]
						
						\[
						5(V_2 - V_1) + V_2 + 10V_2 = 4(V_1 - 12.3)
						\]
						
						\[
						16V_2 - 5V_1 = 4V_1 - 49.2
						\]
					\end{minipage}}

				\noindent\hspace*{38pt}
				\colorbox{yellow}{
					\begin{minipage}{0.6\textwidth}		
						\[
						16V_2 - 9V_1 = -49.2
						\]
						
						\[
						9V_1 - 16V_2 = 49.2 \quad \text{(Equation 2)}
						\]
					\end{minipage}}
			\end{itemize}
			
			\item [4.] \colorbox{yellow}{Solve the System of Linear Equations:}
			\begin{itemize} 
				\item \colorbox{yellow}{We now have two equations:}\\
				
				\noindent\hspace*{38pt}
				\colorbox{yellow}{
					\begin{minipage}{0.6\textwidth}
						\[
						7V_1 - 5V_2 = 24.6 \quad \text{(Equation 1)}
						\]
						
						\[
						9V_1 - 16V_2 = 49.2 \quad \text{(Equation 2)}
						\]
				\end{minipage}}
			\end{itemize}
		\end{itemize}
		~
		
		\colorbox{yellow}{Let's solve these equations for \( V_1 \) and \( V_2 \):}\\
		
		\colorbox{yellow}{From Equation 1,}\\
		
		\noindent\hspace*{70pt}
		\colorbox{yellow}{
			\begin{minipage}{0.6\textwidth}
				\[
				V_2 = \frac{7V_1 - 24.6}{5}
				\]
		\end{minipage}}
		~
		
		\colorbox{yellow}{Substitute \( V_2 \) into Equation 2:}\\
		
		\noindent\hspace*{70pt}
		\colorbox{yellow}{
			\begin{minipage}{0.6\textwidth}
				\[
				9V_1 - 16 \left(\frac{7V_1 - 24.6}{5}\right) = 49.2
				\]
				
				\[
				9V_1 - \frac{112V_1 - 393.6}{5} = 49.2
				\]
				
				\[
				45V_1 - 112V_1 + 393.6 = 246
				\]
				
				\[
				-67V_1 = -147.6
				\]
				
				\[
				V_1 = \frac{147.6}{67} \approx \colorbox{red}{$2.204$}
				\]
		\end{minipage}}
		~
		
		\colorbox{yellow}{Now, substitute \( V_1 \) back into Equation 1:}\\
		
		\noindent\hspace*{70pt}
		\colorbox{yellow}{
			\begin{minipage}{0.6\textwidth}
				\[
				V_2 = \frac{7(2.204) - 24.6}{5} \approx \colorbox{red}{$-1.676$}
				\]
		\end{minipage}}
		~
		
		\colorbox{yellow}{Finally, we find \( i_x \):}\\
		
		\noindent\hspace*{70pt}
		\colorbox{yellow}{
			\begin{minipage}{0.6\textwidth}
				\[
				i_x = \frac{V_1 - 12.3}{10} = \frac{2.204 - 12.3}{10} \approx -1.01 \text{ A}
				\]
		\end{minipage}}
		~
		
		\colorbox{yellow}{Thus, \( i_x \) is approximately \(-1.01\) A.}
	\end{tcolorbox}
\end{center}

This example demonstrates that GPT-4o cannot solve a provided circuit analysis problem, which is expected since GPT-4o cannot even understand the associated circuit diagram. However, we note that GPT-4o does recognize that KCL should be applied to different nodes.

Due to the limitations of current LLMs/VLMs in circuit diagram recognition, we do not include any circuit diagrams in either the reference or student solutions when prompting an LLM to assess homework in this paper. The results in Section \ref{S4} demonstrate the relative effectiveness of our method without image inputs for both GPT-4o and Llama 3 70B.

\subsection{Handwritten and Typed Homework Solution Recognitions}
Other than circuit diagrams, it also appears to be a challenge for VLMs or multimodal language models like GPT-4o to reliably recognize handwritten and typed homework solutions. As shown in the examples in this section, there are errors when we simply prompt GPT-4o to extract the text in the attached image, either in handwritten or typed format, and print it literally in the responses.

The following example shows that the responses of GPT-4o include incorrect information when asked to extract the information from a \textit{handwritten} homework solution\footnote{This homework solution in this example was handwritten by Liangliang Chen, the first author of this paper.} and print it literally. This implies that we cannot use images of student handwritten solutions as direct inputs to GPT-4o for homework assessment. We also demonstrate the results recognized by the Mathpix snipping tool, which is better than that generated by GPT-4o.

\vspace{0.2cm}
\noindent
\img{figures/user_logo.png} User

\begin{center}
	\begin{tcolorbox}[breakable, enhanced, width=\textwidth, opacityfill=1., colback=white]
		\begin{center}
			\frame{\includegraphics[width=0.9\textwidth]{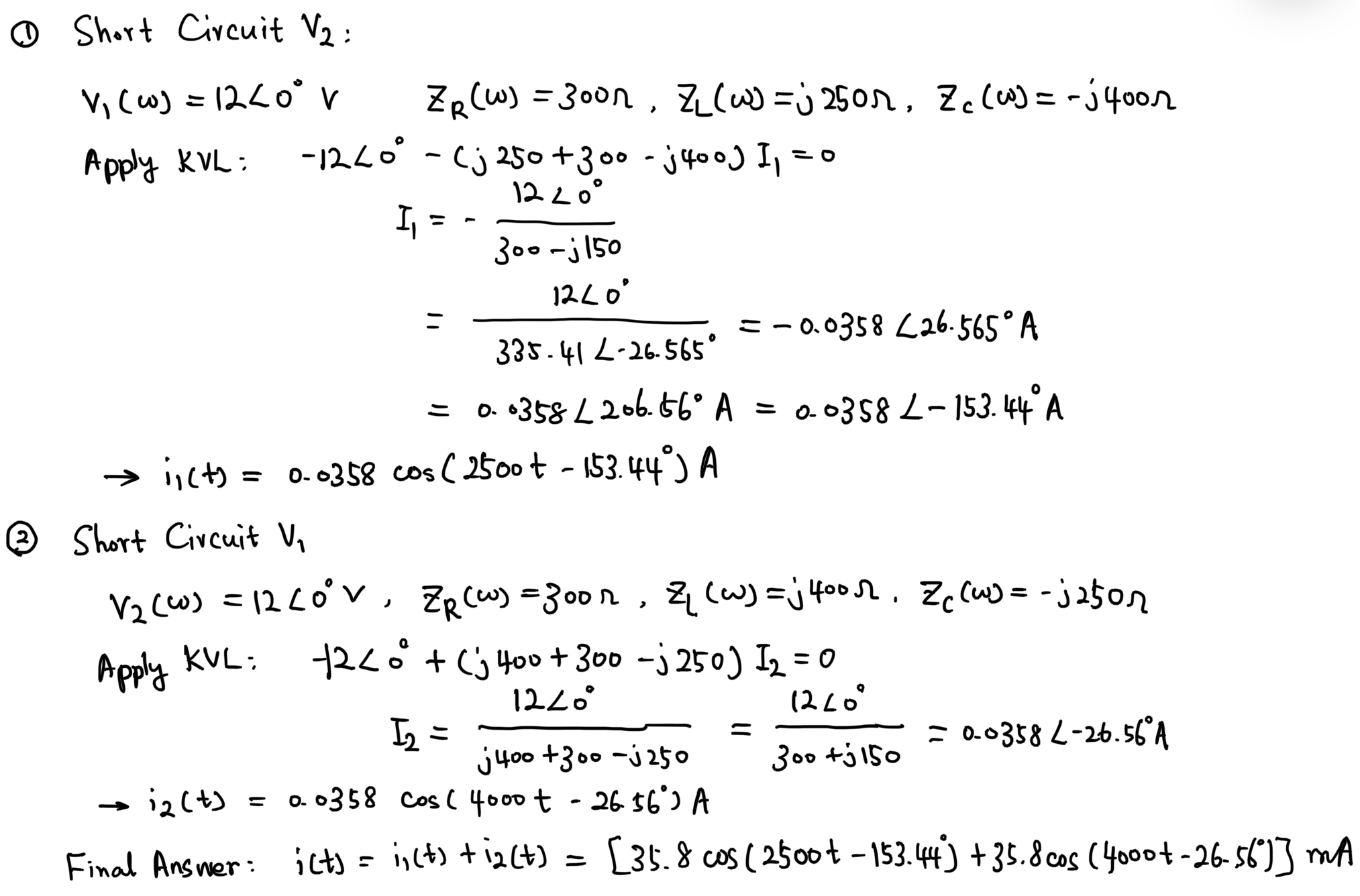}}
		\end{center}
		
		Can you extract the information in the attached image and print it literally in the response? If you encounter equations, you may use LaTeX.
	\end{tcolorbox}
\end{center}

\vspace{0.2cm}
\hfill GPT-4o \img{figures/gpt_logo.png}
\begin{center}
	\begin{tcolorbox}[breakable, enhanced, width=\textwidth, opacityfill=1., colback=white]
		Sure, here is the extracted information from the image:\\
		
		\ding{172} Short Circuit \( V_2 \):\\
		
		\( V_1(\colorbox{red}{$j\omega$}) = 12 \angle 0^\circ \) V\\

		\( Z_R(\colorbox{red}{$j\omega$}) = 300 \Omega \), \( Z_L(\colorbox{red}{$j\omega$}) = j250 \Omega \), \( Z_C(\colorbox{red}{$j\omega$}) = -j400 \Omega \)\\
		
		Apply KVL:
		\[ -12 \angle 0^\circ \colorbox{red}{$- j250 I_1 + 300 I_1 - j400 I_1$} = 0 \]
		
		\[ I_1 = -\frac{12 \angle 0^\circ}{300 - j150} \]
		
		\[ I_1 = \frac{12 \angle 0^\circ}{335.41 \angle -26.565^\circ} = -0.0358 \angle 26.565^\circ \]
		
		\[ I_1 = 0.0358 \angle 206.565^\circ \text{ A} = 0.0358 \angle -153.44^\circ \text{ A} \]
		
		\[ \Rightarrow i_1(t) = 0.0358 \cos(2500 t - 153.44^\circ) \text{ A} \]\\
		
		\ding{173} Short Circuit \( V_1 \):\\
		
		\( V_2(\colorbox{red}{$j\omega$}) = 12 \angle 0^\circ \text{ V} \)\\
		
		\( Z_R(\colorbox{red}{$j\omega$}) = 300 \Omega \), \( Z_L(\colorbox{red}{$j\omega$}) = j400 \Omega \), \( Z_C(\colorbox{red}{$j\omega$}) = -j250 \Omega \)\\
		
		Apply KVL:
		\[ 12 \angle 0^\circ + j400 I_2 + 300 I_2 - j250 I_2 = 0 \]
		
		\[ I_2 = \frac{12 \angle 0^\circ}{j400 + 300 - j250} = \frac{12 \angle 0^\circ}{300 + j150} \]
		
		\[ I_2 = 0.0358 \angle -26.565^\circ \text{ A} \]
		
		\[ \Rightarrow i_2(t) = 0.0358 \cos(4000 t - 26.565^\circ) \text{ A} \]
		~
		
		Final Answer: 
		\[ i(t) = i_1(t) + i_2(t) = [35.8 \cos(2500 t - 153.44^\circ) + 35.8 \cos(4000 t - 26.56^\circ)] \text{ mA} \]
		~
		
		If you need any further assistance or additional details, feel free to ask!
	\end{tcolorbox}
\end{center}

The following example shows that GPT-4o may provide incorrect information when prompted to extract and print the information from a \textit{typed} homework solution.

\vspace{0.2cm}
\noindent
\img{figures/user_logo.png} User

\begin{center}
	\begin{tcolorbox}[breakable, enhanced, width=\textwidth, opacityfill=1., colback=white]
		\begin{center}
			\frame{\includegraphics[width=0.9\textwidth]{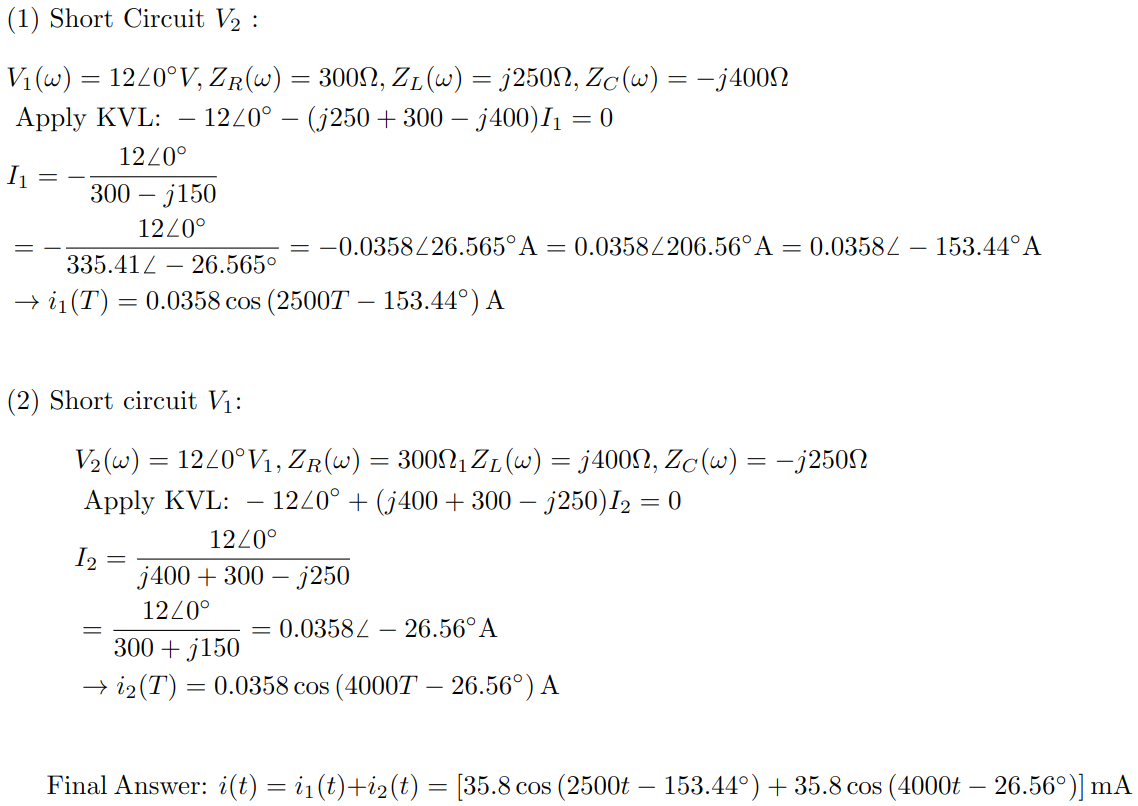}}
		\end{center}
		
		Can you extract the information in the attached image and print it literally in the response? If you encounter equations, you may use LaTeX.
	\end{tcolorbox}
\end{center}

\vspace{0.2cm}
\hfill GPT-4o \img{figures/gpt_logo.png}
\begin{center}
	\begin{tcolorbox}[breakable, enhanced, width=\textwidth, opacityfill=1., colback=white]
		Sure, here is the extracted information from the image:\\
		
		(1) Short Circuit \( V_2 \):
		
		\[ V_1(\omega) = 12 \angle 0^\circ V, Z_R(\omega) = 300 \Omega, Z_L(\omega) = j250 \Omega, Z_C(\omega) = -j400 \Omega \]
		~

		Apply KVL: 

		\[ -12 \angle 0^\circ - (j250 + 300 - j400)I_1 = 0 \]

		\begin{eqnarray*}
			I_1 &=& \colorbox{red}{$\frac{12 \angle 0^\circ}{300 - j150} = \frac{12 \angle 0^\circ}{335.41 \angle -26.565^\circ}$}\\
			&=& -0.0358 \angle 26.565^\circ A = 0.0358 \angle 206.56^\circ A = 0.0358 \angle -153.44^\circ A
		\end{eqnarray*}
		\[ \rightarrow i_1(T) = 0.0358 \cos(2500T - 153.44^\circ) A \]
		~
		
		(2) Short circuit \( V_1 \):
		
		\[ V_2(\omega) = 12 \angle 0^\circ V, Z_R(\omega) = 300 \Omega, Z_L(\omega) = j400 \Omega, Z_C(\omega) = -j250 \Omega \]
		~
		
		Apply KVL:
		\[ -12 \angle 0^\circ + (j400 + 300 - j250)I_2 = 0 \]
		
		\[ I_2 = \frac{12 \angle 0^\circ}{j400 + 300 - j250} = \frac{12 \angle 0^\circ}{300 + j150} = 0.0358 \angle -26.56^\circ A \]
		
		\[ \rightarrow i_2(T) = 0.0358 \cos(4000T - 26.56^\circ) A \]
		~
		
		Final Answer:
		\[ i(t) = i_1(t) + i_2(t) = [35.8 \cos (2500t - 153.44^\circ) + 35.8 \cos (4000t - 26.56^\circ)] mA \]
	\end{tcolorbox}
\end{center}

To demonstrate the performance of the Mathpix snipping tool in converting handwritten solutions, we present the tool’s output for the same example as follows. Compared to the results generated by GPT-4o, the contents extracted by the Mathpix snipping tool contain no significant errors, unlike those in GPT-4o's output. However, the Mathpix snipping tool occasionally produces minor ambiguities. For instance, it sometimes fails to reliably distinguish between uppercase and lowercase letters for certain variables. In the highlighted example, the variables $v_{1}$, $z_{R}$, $z_{L}$, and $z_{c}$ should actually be $V_{1}$, $Z_{\rm R}$, $Z_{\rm L}$, and $Z_{\rm C}$, respectively.\footnote{Since the handwritten solution was prepared by one of the authors, we know that their intended notations are $V_{1}$, $Z_{\rm R}$, $Z_{\rm L}$, and $Z_{\rm C}$.} However, for most cases in students' solutions, it is difficult to determine if a student intended, for example, $z_{c}$ or $Z_{C}$. Therefore, we did not manually correct such minor ``errors'' in our dataset. 

\vspace{0.25cm}
\hspace{-0.5cm}\img{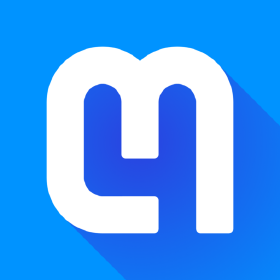} Mathpix Snipping tool
\begin{center}
	\begin{tcolorbox}[breakable, enhanced, width=\textwidth, opacityfill=1., colback=white]
		(1) Short Circuit $V_2$:
		$$
		\colorbox{yellow}{$v_1$}(\omega)=12 \angle 0^{\circ} V \quad \colorbox{yellow}{$z_R$}(\omega)=300 \Omega, \colorbox{yellow}{$z_L$}(\omega)=j 250 \Omega, \colorbox{yellow}{$z_c$}(\omega)=-j 400 \Omega
		$$
		
		Apply KVL: $\quad-12 \angle 0^{\circ}-(j 250+300-j 400) I_1=0$
		
		$$
		\begin{aligned}
			& I_1=-\frac{12 \angle 0^{\circ}}{300-j 150} \\
			&=\frac{12 \angle 0^{\circ}}{335.41 \angle-26.565^{\circ}}=-0.0358 \angle 26.565^{\circ} \mathrm{A} \\
			&=0.0358 \angle 206.56^{\circ} \mathrm{A}=0.0358 \angle-153.44^{\circ} \mathrm{A} \\
			& \rightarrow i_1(t)=0.0358 \cos \left(2500 t-153.44^{\circ}\right) \mathrm{A}
		\end{aligned}
		$$
		
		(2) Short Circuit $V_1$
		
		$$
		V_2(\omega)=12 \angle 0^{\circ} V, \colorbox{yellow}{$z_R$}(\omega)=300 \Omega, \colorbox{yellow}{$z_L$}(\omega)=j 400 \Omega, \colorbox{yellow}{$z_c$}(\omega)=-j 250 \Omega
		$$
		
		Apply KVL: $\quad-12 \angle 0^{\circ}+(j 400+300-j 250) I_2=0$
		
		$$
		\begin{aligned}
			& I_2=\frac{12 \angle 0^{\circ}}{j 400+300-j 250}=\frac{12 \angle 0^{\circ}}{300+j 150}=0.0358 \angle-26.56^{\circ} \mathrm{A} \\
			& \rightarrow i_2(t)=0.0358 \cos \left(4000 t-26.56^{\circ}\right) \mathrm{A}
		\end{aligned}
		$$

		Final Answer: $\quad i(t)=i_1(t)+i_2(t)=[35.8 \cos \left(2500 t-153.44^{\circ}\right)+35.8 \cos \left(4000 t-26.56^{\circ}\right)] \mathrm{mA}$
	\end{tcolorbox}
\end{center}

It is also important to note that the handwritten solution in this example is relatively neat. In contrast, some students' solutions exhibit more varied and freeform writing styles. For example, we observed the following recognition errors, which we corrected during the proofreading process.

\begin{itemize}
	\item ``$\sqrt{(\omega)}$'' was corrected to ``$V(\omega)$''.
	\item ``$2\cos(-45^{\circ})+j^{2}\sin(-45^{\circ})$'' was corrected to ``$2\cos(-45^{\circ})+j2\sin(-45^{\circ})$''.
	\item ``$\dot{I}$'' was corrected to ``$i$''.
\end{itemize}

As illustrated by the examples in this section, the Mathpix Snipping Tool demonstrates greater reliability in recognizing homework solutions compared to GPT-4o. Consequently, we used the Mathpix Snipping Tool to convert both the reference and student solutions into LaTeX documents. These solutions were then incorporated into the overall structured prompt as input for different LLMs. The results in Section \ref{S4} show that the tested LLMs (GPT-3.5 Turbo, GPT-4o, and Llama 3 70B) successfully interpret LaTeX sources, validating the effectiveness of our alternative method.

\subsection{Mathematical Calculations}
Consistent with the existing work \citep{frieder2023mathematical} that has demonstrated that GPT-4 can offer sufficient (but not perfect) performance for undergraduate-level mathematics, we note that specific mathematical calculations, such as the conversion between phasors and sinusoids, may be challenging for current state-of-the-art LLMs, including GPT-4o.

We provide the following example where intentional modifications were made to the correct solution of P 10.4-6 in \citep{svoboda2013introduction} to create a solution with arithmetic errors. We then tested GPT-4o to see if it could correctly identify the arithmetic errors in the solution independently, without the assistance of the official solution. 

The considered problem is given as follows. 

\vspace{0.1cm}
\begin{center}
	\frame{\includegraphics[width=0.5\textwidth]{figures/Example_2.png}}
\end{center}
\vspace{0.1cm}

The correct solution to this problem is presented as follows.
\begin{tcolorbox}[breakable, enhanced, width=\textwidth, opacityfill=1., colback=lightgray]
	Denote the voltage at the junction of the 10 \(\Omega\), 4 \(\Omega\), and 20 \(\Omega\) resistors as \(v_{1}\), and the voltage at the junction of the 5 \(\Omega\), 20 \(\Omega\), and 2 \(\Omega\) resistors and the current-controlled current source as \(v_{2}\), respectively.
	
	Step 1: According to Ohm's law, we have
	\[
	i_{x}=\frac{12.3-v_{1}}{10}
	\]
	which indicates
	\[
	2i_{x}=\frac{12.3-v_{1}}{5}
	\]
	
	Step 2: By applying Kirchhoff's Current Law (KCL) to node 1, where the voltage is $v_{1}$, we have
	
	\[ 
	\frac{v_{1}-12.3}{10}+\frac{v_{1}}{4}+\frac{v_{1}-v_{2}}{20}=0
	\]
	
	which can be simplified to 
	
	\[
	8v_{1}-v_{2}=24.6
	\]

	Step 3: By applying KCL to node 2, where the voltage is $v_{2}$, we have
	
	\[
	\frac{v_{2}-v_{1}}{20}+\frac{v_{2}-12.3}{5}+\frac{v_{2}}{2}-\frac{12.3-v_{1}}{5}=0
	\]
	which can be simplified to 
	\[
	v_{1}+5v_{2}=32.8
	\]
	
	Solving the system of equations \(8v_{1}-v_{2}=24.6\) and \(v_{1}+5v_{2}=32.8\), we have
	
	\[
	v_{1}=3.8 \text{ V},~v_{2}=5.8 \text{ V}
	\]
	
	Thus, 
	\[
	i_{x}=\frac{12.3-3.8}{10}=0.85\text{ A}
	\]
\end{tcolorbox}

The following box represents an incorrect solution, in which we intentionally introduce an arithmetic error.
\begin{tcolorbox}[breakable, enhanced, width=\textwidth, opacityfill=1., colback=lightgray]
	Denote the voltage at the junction of the 10 \(\Omega\), 4 \(\Omega\), and 20 \(\Omega\) resistors as \(v_{1}\), and the voltage at the junction of the 5 \(\Omega\), 20 \(\Omega\), and 2 \(\Omega\) resistors and the current-controlled current source as \(v_{2}\), respectively.
	
	Step 1: According to Ohm's law, we have
	\[
	i_{x}=\frac{12.3-v_{1}}{10}
	\]
	which indicates
	\[
	2i_{x}=\frac{12.3-v_{1}}{5}
	\]
	
	Step 2: By applying Kirchhoff's Current Law (KCL) to node 1, where the voltage is $v_{1}$, we have
	
	\[ 
	\frac{v_{1}-12.3}{10}+\frac{v_{1}}{4}+\frac{v_{1}-v_{2}}{20}=0
	\]
	
	which can be simplified to 
	
	\[
	8v_{1}-v_{2}=\colorbox{red}{$-24.6$}
	\]

	Step 3: By applying KCL to node 2, where the voltage is $v_{2}$, we have
	
	\[
	\frac{v_{2}-v_{1}}{20}+\frac{v_{2}-12.3}{5}+\frac{v_{2}}{2}-\frac{12.3-v_{1}}{5}=0
	\]
	which can be simplified to 
	\[
	v_{1}+5v_{2}=\colorbox{red}{$-32.8$}
	\]
	
	Solving the system of equations \(8v_{1}-v_{2}=\colorbox{red}{$-24.6$}\) and \(v_{1}+5v_{2}=\colorbox{red}{$-32.8$}\), we have
	
	\[
	\colorbox{red}{$v_{1}=-3.8 \text{ V},~v_{2}=-5.8 \text{ V}$}
	\]
	
	Thus, 
	\[
	\colorbox{red}{$i_{x}=\frac{12.3-(-3.8)}{10}=1.61\text{ A}$}
	\]
\end{tcolorbox}

We prompt GPT-4o to identify the arithmetic error in the incorrect solution provided above. Due to GPT-4o's unreliability of image recognition, we directly use the LaTeX source of the incorrect solution in the prompt. Although GPT-4o correctly derived the intermediate equations $8v_{1}-v_{2}=24.6$ and $v_{1}+5v_{2}=32.8$, it consistently used the wrong equation $8v_{1}-v_{2}=-24.6$ in the following calculation processes, leading to another incorrect answer.

\vspace{0.2cm}
\noindent
\img{figures/user_logo.png} User
\begin{center}
	\begin{tcolorbox}[breakable, enhanced, width=\textwidth, opacityfill=1., colback=white]
		The following solution in LaTeX shows a student solution to a circuit analysis problem. Can you identify if there is any arithmetic error in the solution?\\
		
		Denote the voltage at the junction of the 10 \verb|\(\Omega\)|, 4 \verb|\(\Omega\)|, and 20 \verb|\(\Omega\)| resistors as \verb|\(v_{1}\)|, and the voltage at the junction of the 5 \verb|\(\Omega\)|, 20 \verb|\(\Omega\)|, and 2 \verb|\(\Omega\)| resistors and the current-controlled current source as \verb|\(v_{2}\)|, respectively.\\
		
		Step 1: According to Ohm's law, we have\\
		
		\verb|\[i_{x}=\frac{12.3-v_{1}}{10}\]|
		
		~
		
		which indicates\\
		
		\verb|\[2i_{x}=\frac{12.3-v_{1}}{5}\]|
		
		~

		Step 2: By applying Kirchhoff's Current Law (KCL) to node 1, where the voltage is \verb|\(v_{1}\)|, we have\\
		
		\verb|\[\frac{v_{1}-12.3}{10}+\frac{v_{1}}{4}+\frac{v_{1}-v_{2}}{20}=0\]|
		
		~
		
		which can be simplified to \\
		
		\verb|\[8v_{1}-v_{2}=-24.6\]|
		
		~
		
		Step 3: By applying KCL to node 2, where the voltage is \verb|\(v_{2}\)|, we have\\
		
		\verb|\[\frac{v_{2}-v_{1}}{20}+\frac{v_{2}-12.3}{5}+\frac{v_{2}}{2}\]|
		
		\verb|-\frac{12.3-v_{1}}{5}=0\]|
		
		~
		
		which can be simplified to \\
		
		\verb|\[v_{1}+5v_{2}=-32.8\]|
		
		~
		
		Solving the system of equations \verb|\(8v_{1}-v_{2}=-24.6\)| and \verb|\(v_{1}+5v_{2}=| \verb|-32.8\)|, we have\\
		
		\verb|\[v_{1}=-3.8 \text{ V},~v_{2}=-5.8 \text{ V}\]|
		
		~
		
		Thus,\\

		\verb|\[i_{x}=\frac{12.3-(-3.8)}{10}=1.61\text{ A}\]|
	\end{tcolorbox}
\end{center}

\vspace{0.2cm}
\hfill GPT-4o \img{figures/gpt_logo.png}
\begin{center}
	\begin{tcolorbox}[breakable, enhanced, width=\textwidth, opacityfill=1., colback=white]
		Let's go through the solution step by step and check for arithmetic errors.\\
		
		\textbf{Step 1: Current \(i_x\)}\\
		
		Given:
		\[ i_{x} = \frac{12.3 - v_{1}}{10} \]
		
		\[ 2i_{x} = \frac{12.3 - v_{1}}{5} \]
		~
		
		These are consistent with Ohm's law and the given problem statement.\\
		
		\textbf{Step 2: KCL at Node 1}\\
		
		Applying KCL at node 1 (\(v_{1}\)):
		
		\[ \frac{v_{1} - 12.3}{10} + \frac{v_{1}}{4} + \frac{v_{1} - v_{2}}{20} = 0 \]
		~
		
		Combining the fractions over a common denominator:
		
		\[ \frac{2(v_{1} - 12.3) + 5v_{1} + (v_{1} - v_{2})}{20} = 0 \]
		
		\[ 2v_{1} - 24.6 + 5v_{1} + v_{1} - v_{2} = 0 \]
		\[\colorbox{green}{$8v_{1} - v_{2} = 24.6$}\]
		~
		
		It should be:
		
		\[ \colorbox{red}{$8v_{1} - v_{2} = -24.6$} \]
		~
		
		\colorbox{red}{This step seems to be correct.}\\
		
		\textbf{Step 3: KCL at Node 2}\\
		
		Applying KCL at node 2 (\(v_{2}\)):
		
		\[ \frac{v_{2} - v_{1}}{20} + \frac{v_{2} - 12.3}{5} + \frac{v_{2}}{2} - \frac{12.3 - v_{1}}{5} = 0 \]
		~
		
		Combining the fractions over a common denominator:
		
		\[ \frac{v_{2} - v_{1}}{20} + \frac{v_{2} - 12.3}{5} + \frac{v_{2}}{2} - \frac{12.3 - v_{1}}{5} \]
		~
		
		Simplifying each term:
		
		\begin{align*}
			&\frac{v_{2} - v_{1}}{20} + \frac{v_{2} - 12.3}{5} + \frac{v_{2}}{2} - \frac{12.3 - v_{1}}{5}\\
			=~&\frac{v_{2} - v_{1}}{20} + \frac{v_{2} - 12.3}{5} + \frac{v_{2}}{2} - \frac{12.3 - v_{1}}{5}\\
			=~&\frac{v_{2} - v_{1}}{20} + \frac{v_{2} - 12.3 + v_{1} - 12.3}{5} + \frac{v_{2}}{2}\\
			=~&\frac{v_{2} - v_{1}}{20} + \frac{v_{2} + v_{1} - 24.6}{5} + \frac{v_{2}}{2}
		\end{align*}
		~
		
		Combining terms over a common denominator:
		
		\[ \frac{v_{2} - v_{1}}{20} + 4 \left( \frac{v_{2} + v_{1} - 24.6}{20} \right) + 10 \left( \frac{v_{2}}{20} \right) = 0 \]
		\[ \frac{v_{2} - v_{1} + 4v_{2} + 4v_{1} - 98.4 + 10v_{2}}{20} = 0 \]
		\[ \frac{15v_{2} + 3v_{1} - 98.4}{20} = 0 \]
		\[ 15v_{2} + 3v_{1} - 98.4 = 0 \]
		\[ 3v_{1} + 15v_{2} = 98.4 \]
		\[ v_{1} + 5v_{2} = 32.8 \]
		~
		
		This equation should be:
		
		\[ \colorbox{green}{$v_{1} + 5v_{2} = 32.8$} \]
		~
		
		\textbf{Solving the system of equations:}\\
		
		The system of equations is:\\
		
		1. \( \colorbox{red}{$8v_{1} - v_{2} = -24.6$} \)
		
		2. \( \colorbox{green}{$v_{1} + 5v_{2} = 32.8$} \)
		
		~
		
		Solving these simultaneously:\\
		
		Multiply the second equation by 8 to eliminate \(v_{1}\):
		
		\[ 8v_{1} + 40v_{2} = 262.4 \]
		~
		
		Subtract the first equation from this result:
		\[ (8v_{1} + 40v_{2}) - (8v_{1} - v_{2}) = 262.4 \colorbox{red}{$+ 24.6$} \]
		
		\noindent\hspace*{-8pt}
		\colorbox{yellow}{
			\begin{minipage}{1.01\textwidth}
				\[ 41v_{2} = 287 \]
				\[ v_{2} = \frac{287}{41} \]
				\[ v_{2} = 7 \]
				
				Substituting \(v_{2} = 7\) into \( v_{1} + 5v_{2} = 32.8 \):
				
				\[ v_{1} + 35 = 32.8 \]
				\[ v_{1} = 32.8 - 35 \]
				\[ v_{1} = -2.2 \]
				~
				
				Thus:
				\[ v_{1} = -2.2 \text{ V}, \]
				\[ v_{2} = 7 \text{ V} \]
				~
				
				Thus,
				\[ i_{x} = \frac{12.3 - (-2.2)}{10} \]
				\[ i_{x} = \frac{12.3 + 2.2}{10} \]
				\[ i_{x} = \frac{14.5}{10} \]
				\[ i_{x} = 1.45 \text{ A} \]
				~
				
				So, there was an arithmetic error in solving the system of equations. The corrected values should be \( v_{1} = -2.2 \) V and \( v_{2} = 7 \) V, leading to \( i_{x} = 1.45 \) A, not 1.61 A.
		\end{minipage}}
	\end{tcolorbox}
\end{center}

The results show that GPT-4o failed to point out the arithmetic errors and incorrectly stated that the solution was correct. Thus, there are still some limitations for GPT-4o in evaluating student homework solutions independently. This limitation motivates us to include the LaTeX document of the official solution as a part of the LLM prompt in the method used in this paper.

\section{Discussions and Future Directions}
This section presents the discussions of this paper and some future directions. 

\subsection{Reliability Improvement by Prompt Engineering and LLM Fine-tuning}
According to the results in Table \ref{Tab0} of Section \ref{S4}, LLMs, particularly GPT-4o, exhibit promising capabilities in homework assessment for the circuit analysis course. However, due to the critical need for reliability when providing assessment results to students, it is essential to explore strategies to enhance the reliability of LLMs through prompt engineering and model fine-tuning. In this paper, a common prompt template was used for all three evaluated LLMs to ensure fair comparisons of their responses. In future research, we plan to develop customized prompts tailored to each LLM to further optimize their performance in assessing homework. The limitations identified in this study provide valuable insights for refining prompt designs or fine-tuning LLMs with appropriate data. We are actively investigating approaches to improve off-the-shelf models for homework assessment in circuit analysis courses.

\subsection{Information Extraction from Circuit Diagrams}
In this paper, the official reference solution serves as a benchmark for student homework assessment. It also provides essential information for LLMs in conducting this assessment process. Circuit diagrams, however, are crucial elements in circuit analysis courses. Although the state-of-the-art model, GPT-4o, cannot yet reliably recognize elements in a basic circuit diagram, we can design and employ external tools to extract information from these diagrams and present it in a way that LLMs can interpret. This approach will enable LLMs to conduct assessment even when no reference solution is available. Future research can further explore the use of LLMs to independently solve circuit analysis problems. Additionally, it would be also worthwhile to investigate how to appropriately represent and process extracted information from circuit diagrams to enhance LLMs' comprehension \citep{hakhamaneshi2022pretraining, yamakaji2024circuit2graph, jin2024large}.

\subsection{Generalizability of the Evaluation Framework}
The dataset used in this paper is drawn solely from the circuit analysis course, and accordingly, the LLMs are evaluated only on homework assessment for this subject. In future work, it would be valuable to expand the dataset to include additional courses in electrical and computer engineering, as well as other areas within the broader engineering field. To generalize the evaluation framework to other domains, we may consider the intrinsic capabilities of LLMs, the construction of the dataset, and the selection of appropriate metrics. 

\begin{itemize}
\item If the LLMs to be evaluated are sufficiently powerful, they should ideally be able to generate reasonable homework assessments based solely on their intrinsic capabilities. Thus, when adapting the evaluation framework in this paper to other domains, it is advisable to first test a few examples to determine whether the evaluated LLMs possess adequate intrinsic capabilities for homework assessment without the support of additional datasets. If not, it would be necessary to evaluate the LLMs' performance by augmenting the prompts with appropriate external data, such as reference solutions.

\item To prepare evaluation data, it is important to consider whether diagrams are essential components of the problem descriptions and solutions. If they are, it is necessary to first assess whether LLMs can interpret these diagrams or whether alternative sources of information can be used to prompt the LLMs. In other words, datasets containing the necessary information for homework evaluation should be constructed to effectively leverage the evaluation framework. To further ensure \textit{reliable} homework assessment, it is crucial that the dataset provides sufficient information that can be readily understood by LLMs, either directly or through appropriate adaptation.

\item From a metric perspective, the evaluation criteria should be adapted to different domains when applying the proposed framework. In Section \ref{S32} of this paper, we selected five metrics---completeness, method, final answer, arithmetic, and unit---to evaluate LLM performance within the domain of undergraduate-level circuit analysis. However, these metrics should be modified to suit the specific contexts of other domains. Several existing studies have explored LLMs' assessment capabilities across various fields, which may serve as useful references for researchers and practitioners in other areas.

For instance, \cite{zhu2024impact} assessed the assignment completion ability of LLMs in the context of middle school students’ learning, using a scoring rubric that included criteria such as completion, creativity, accuracy, material refinement, depth, logic, and page aesthetics. Similarly, \cite{chiang2023can} evaluated LLM-generated open-ended stories based on grammaticality, cohesiveness, likability, and relevance. In another example, \cite{wang2024enhancing} examined LLM performance in Python code generation using various prompt-engineering strategies, with evaluations focusing on metrics such as pass rate, time spent, and Pylint score. In summary, the choice of evaluation metrics and associated prompt designs is task-specific and remains flexible in the evolving landscape of LLM development.
\end{itemize} 

\section{Conclusion}
\label{S6}
In this paper, we benchmark three LLMs, specifically GPT-3.5 Turbo, GPT-4o, and Llama 3 70B, on homework assessment for an undergraduate-level circuit analysis course. The LLMs are evaluated using five different metrics on our introduced dataset composed of real student solutions and official solutions. The results demonstrate that GPT-4o and Llama 3 70B can generate relatively reasonable homework assessments, with their reliability varying according to the difficulty of different topics. We summarize the strengths and weaknesses of GPT-4o and Llama 3 70B. Our experiments also reveal GPT-4o's limitations in understanding circuit diagrams, recognizing handwritten and typed homework solutions, and performing independent mathematical calculations. Our future work will investigate how to build a personalized tutor in circuit analysis by leveraging the advantages of LLMs and circumventing their weaknesses identified in this paper.

\begin{appendices}
	
\section{Dataset Details}
\label{SA}
\begin{center}
	\begin{figure}[!h]
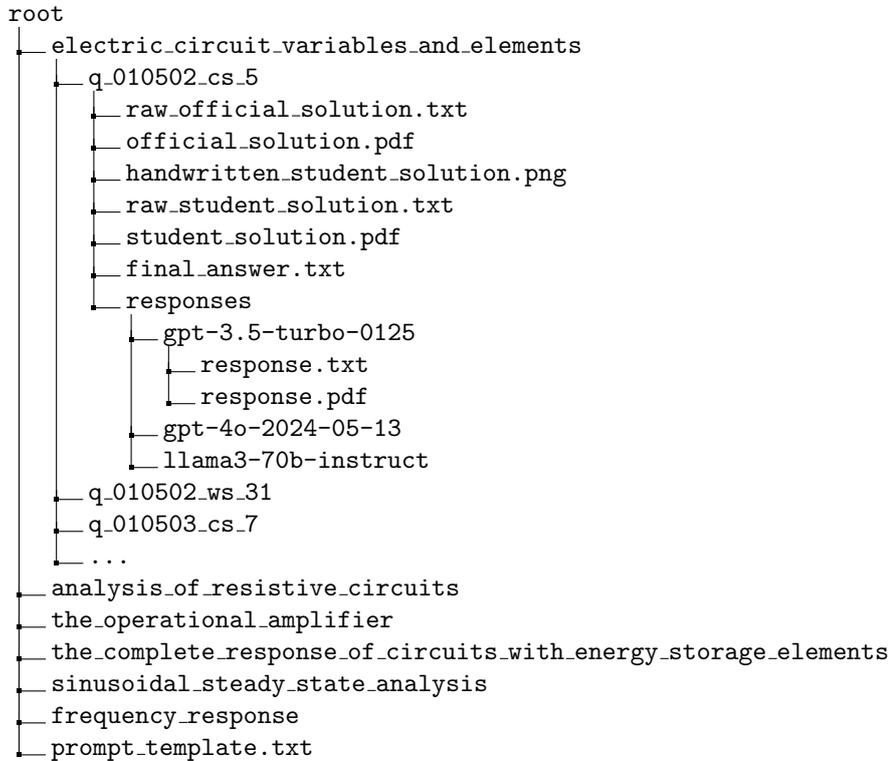

		\raggedright
		\dirtree{%
			.1 root.
			.2 electric\_circuit\_variables\_and\_elements.
			.3 q\_010502\_cs\_5.
			.4 raw\_official\_solution.txt.
			.4 official\_solution.pdf.
			.4 handwritten\_student\_solution.png.
			.4 raw\_student\_solution.txt.
			.4 student\_solution.pdf.
			.4 final\_answer.txt.
			.4 responses.
			.5 gpt-3.5-turbo-0125.
			.6 response.txt.
			.6 response.pdf.
			.5 gpt-4o-2024-05-13.
			.5 llama3-70b-instruct.
			.3 q\_010502\_ws\_31.
			.3 q\_010503\_cs\_7.
			.3 .... 
			.2 analysis\_of\_resistive\_circuits.
			.2 the\_operational\_amplifier.
			.2 the\_complete\_response\_of\_circuits\_with\_energy\_storage\_elements.
			.2 sinusoidal\_steady\_state\_analysis.
			.2 frequency\_response.
			.2 prompt\_template.txt.
		}
		\caption{Directory tree of the dataset}
		\label{FigA1}
	\end{figure}
\end{center}

This document details the dataset collected and used in this paper. The directory tree of the dataset is shown in Fig. \ref{FigA1}. 

Note that the dataset used in this paper will not be shared with the community, as the students enrolled in the Fall 2021 circuit analysis course did not sign a consent form, and their solutions are integral to this dataset. However, we are actively collecting additional student homework data from various courses, including circuit analysis. In future publications, we plan to release datasets collected from students who have provided consent.

The basic elements in the dataset are structured based on different students' solutions to the associated problems. The problems in our dataset are all from the textbook \citep{svoboda2013introduction}, and the students' solutions are collected from a circuit analysis course for undergraduates involving 37 students in the School of Electrical and Computer Engineering at the Georgia Institute of Technology. The folders of these basic elements are named with the following method, in which we take \texttt{q\_010502\_cs\_5} as an example to illustrate.  

\begin{itemize}
	\item [i)] In this example, ``\texttt{q\_010502}'' means that the corresponding problem is Problem 1.5-2 in \citep{svoboda2013introduction}. Specifically, the codes ``\texttt{01}'' and ``\texttt{05}'' indicate the chapter and section indexes, respectively, while ``\texttt{02}'' is the problem index in the section.
	\item [ii)] The code ``\texttt{cs\_5}'' indicates that the included student's solution in the folder \texttt{q\_010502\_cs\_5} is from the student with index \texttt{5}. The student indices are mainly for our internal use to track the original documents of students' solutions. They are de-identified and consistently used across different problems. Additionally, ``\texttt{cs}'' implies that student \texttt{5}'s solution is considered correct in all metrics. We use the code ``\texttt{ws}'' to represent the students' solutions that have errors in at least one metric out of the five considered. For example, ``\texttt{q\_010502\_ws\_31}'' indicates that the solution from student \texttt{31} solves Problem 1.5-2 in \citep{svoboda2013introduction} with at least one error in one aspect.
\end{itemize}

The following data are included in the basic data folder like \texttt{q\_010502\_cs\_5}.
\begin{itemize}
	\item \texttt{raw\_official\_solution.txt}: This file contains the official solution to the corresponding problem in LaTeX format. The contents of this file are inserted in the LLM prompt when we leverage LLMs to assess students' homework. 
	
	\item \texttt{official\_solution.pdf}: For a better reading experience, we generated a PDF file corresponding to the official solution \texttt{raw\_official\_solution.txt}.
	
	\item \texttt{handwritten\_student\_solution.png}: This is the original source file of the student's solution. To test the capabilities of LLMs in homework assessment, we converted this image to a LaTeX document using the Mathpix snipping tool. The converted results are saved in the file \texttt{raw\_student\_solution.txt}. 
	
	\item \texttt{raw\_student\_solution.txt}: This file contains the student's solution in LaTeX format. The contents of this file are included in the prompt and evaluated by the tested LLMs.
	
	\item \texttt{student\_solution.pdf}: For a better reading experience, we generated a PDF file corresponding to the student's solution \texttt{raw\_student\_solution.txt}. 
	
	\item \texttt{final\_answer.txt}: This file contains a short summary of the final answer to the corresponding problem. Comparing final answers is an easy task for the LLM, and a student's incorrect final answer may motivate the LLM to identify where the student made the error. The contents of this file are also included as part of the LLM prompt. 
	
	The final answer provides a concise summary of the official solution, omitting the problem-solving process. For example, a circuit problem may ask students to determine (a) the current $i_{R}$ through a resistor $R$, (b) the voltage $v_{R}$ across the resistor, and (c) the power $p_{R}$ absorbed by this resistor. The final answer might be presented as: ``(a) $i_{R} = 1~\mathrm{A}$; (b) $v_{R} = 1~\mathrm{V}$; (c) $p_{R} = 1~\mathrm{W}$.'' In contrast, the official solution includes a detailed derivation of these values, showing how relevant circuit laws and analysis methods are applied.
		
	Including the final answer in the prompt helps LLMs avoid overlooking important information when assessing student solutions. This is particularly useful when evaluating metrics such as completeness and accuracy of final answers, especially in problems with multiple sub-parts, like the example in the previous paragraph.
	
	\item \texttt{responses}: This file consists of the responses from the three LLMs evaluated in this paper. The evaluation results are structured into three subfolders: \texttt{gpt-3.5-turbo-0125}, \texttt{gpt-4o-2024-05-13}, and \texttt{llama3-70b-instruct}, according to the models. Each subfolder contains two files: \texttt{response.txt}, which saves the direct LLM output, and \texttt{response.pdf}, which is the corresponding PDF file of \texttt{response.txt} for a better reading experience. The evaluation classification results are obtained by human inspection based on the file \texttt{response.pdf}. 
\end{itemize}

\section{Notable LLM Response Examples}
\label{SD2}
In order to better demonstrate the strengths and weaknesses of different LLMs, especially GPT-4o and Llama 3 70B, this document provides notable examples of LLM responses in different topics. We use ``{\color{green}\ding{168}}'' to denote the examples that can reflect the strengths of LLMs and ``{\color{red}\ding{171}}'' to denote those that reflect the weaknesses.

\subsection{Electric Circuit Variables and Elements}
\label{SD21}
\noindent
\textbf{GPT-4o:}
\begin{itemize}
	\item [{\color{red}\ding{171}}] \texttt{q\_010502\_cs\_7} [Final Answer]: The correct cost should be \$0.264, but your final answer is 26.4 cents (i.e., \$0.264). It seems you made an error when interpreting units (hours should have been used instead of seconds).
	
	\textit{Note:} GPT-4o fails to identify the equivalence between 26.4 cents and \$0.264.
	
	\item [{\color{green}\ding{168}}] \texttt{q\_010503\_cs\_35} [Final Answer]: The student's final answer of $3{,}333.33~\text{s}$ is equivalent to the official solution's $3.33 \times 10^3~\text{s}$. This confirms that the answer is correct when considering significant figures and notational differences.
	
	\textit{Note:} GPT-4o can recognize that $3{,}333.33~\text{s}$ is equivalent to $3.33 \times 10^3~\text{s}$.
	
	\item [{\color{green}\ding{168}}] \texttt{q\_010503\_cs\_37} [Unit]: Your notation is slightly different but acceptable ($p$ instead of $P$), as you were consistent with it.
	
	\textit{Note:} GPT-4o can accommodate different notation in the student's solution as long as the notation is used consistently. 
	
	\item [{\color{green}\ding{168}}] \texttt{q\_020406\_cs\_6} [Unit]: When calculating the power absorbed by \( R_1 \), the student writes the intermediate result as \(0.0025~ \mathrm{\omega}\) which is incorrect. It should be \(0.0025 \text{~W}\) instead. This typographical error has been correctly converted to \(2.5 \text{~mW}\) in the final step.
	
	\textit{Note:} GPT-4o points out the typo of the unit $\omega$, which should be W, demonstrating its robustness to occasional typos. 
	
	\item [{\color{green}\ding{168}}] \texttt{q\_020406\_ws\_7} [Final Answer]:
	\begin{itemize}
		\item [--] For \(v_1\), the student calculated \(-100 \, \text{mV}\), which matches the correct answer \(-0.1 \, \text{V}\). The conversion between volts and millivolts is appropriate.
		\item [--] For \(v_2\), the student calculated \(200 \, \text{mV}\), which matches the correct answer \(0.2 \, \text{V}\).
		
		\textit{Note:} GPT-4o recognizes the equivalences of answers in different units. 
	\end{itemize}
	
	\item [{\color{red}\ding{171}}] \texttt{q\_020406\_ws\_9} [Completeness]: The student's solution is incomplete. While they calculated the voltage across the resistors and the power absorbed by each resistor, they did not address the direction of current and adherence to the passive sign convention.
	
	\textit{Note:} The student's solution should be considered complete since they calculated the voltage across the resistors and the power absorbed by each resistor.
	
	\item [{\color{red}\ding{171}}] \texttt{q\_020406\_ws\_18} [Unit]: The units for power are inconsistent and incorrect due to the sign convention error. Additionally, the answer is expected in milliwatts (mW), but the unit was given in watts (W) which adds to the confusion. 
	
	\textit{Note:} GPT-4o is overly confident that the answer is expected in milliwatts. In fact, the answer given in watts is also acceptable.
	
	\item [{\color{green}\ding{168}}] \texttt{q\_020501\_cs\_6} [Unit]: In part (b), there's a minor typo in the student's solution ``15 vV'' which should be ``15 V''. This is a small error and doesn't affect the correctness of the calculation but should be noted for precision and clarity.
	
	\textit{Note:} GPT-4o can recognize the unit typo and state that the small typo does not affect the correctness of the solution.
	
	\item [{\color{green}\ding{168}}] \texttt{q\_020501\_cs\_11} [Method]: The power calculation method the student used \( p = I \cdot V \) is correct but it is different from the official solution's method \( P = R i^2 \). Both methods are valid and yield the correct result.
	
	\textit{Note:} GPT-4o correctly identifies an alternative approach as correct. 
\end{itemize}

\noindent
\textbf{Llama 3 70B:}
\begin{itemize}
	\item [{\color{red}\ding{171}}] \texttt{q\_010502\_cs\_7} [Final Answer]: The student's final answers are mostly correct, but with some errors:
	\begin{itemize}
		\item [--] Part (a): The student's answer, $q = 72000 \mathrm{C}$, is incorrect. The correct answer is $q = 7.2 \times 10^4 \mathrm{C}$. The student's mistake is likely due to a typo or a misunderstanding of the unit conversion.
		\item [--] Part (b): The student's answer, $P = 1100 \mathrm{~W}$, is correct.
		\item [--] Part (c): The student's answer, $cost = 26.4 \text{ cents}$, is incorrect. The correct answer is $cost = \$0.264$. The student's mistake is due to incorrect unit conversion and cost calculation.
	\end{itemize}
	
	\textit{Note:} Llama 3 70B does not recognize the equivalence between 72000 and $7.2\times10^{4}$ as well as between 26.4 cents and \$0.264.
	
	\item [{\color{red}\ding{171}}] \texttt{q\_010502\_cs\_27} [Final Answer]:
	\begin{itemize}
		\item [--] Part (a): The student's answer, $q=72,000 \mathrm{C}$, is close to the correct answer, $q=7.2 \times 10^4 \mathrm{C}$. The student's answer is in the correct units (Coulombs) and is approximately correct, but not exactly equal to the correct answer. This might be due to a rounding error or a slight calculation mistake.
		\item [--] Part (b): The student's answer, $P=1100 \mathrm{~W}$, is correct.
		\item [--] Part (c): The student's answer, $\approx 264$, is close to the correct answer, $\$0.264$. The student's answer is in the correct units (dollars) and is approximately correct, but not exactly equal to the correct answer. This might be due to a rounding error or a slight calculation mistake.
	\end{itemize}
	\textit{Note:} Llama 3 70B does not recognize that 72,000 C and $7.2\times10^{4}$ C are exactly the same value. The assessment to part (c) is also dubious, as 264 cents are not close to \$0.264.
	
	\item [{\color{green}\ding{168}}] \texttt{q\_010503\_cs\_5} [Final Answer]: The student's final answer, 3333.33 seconds, is correct. It matches the official solution's answer of $\Delta t = 3.33 \times 10^3~\text{s}$.
	
	\textit{Note:} Llama 3 70B successfully recognizes the equivalence between 3333.33 seconds and $3.33\times10^{3}$ seconds.
	
	\item [{\color{red}\ding{171}}] \texttt{q\_020406\_ws\_16} [Arithmetic]: There are arithmetic errors in the student's solution. For example, in calculating $P_1$, the student writes $-0.0025$ W, which is incorrect. The correct calculation is $P_1=-v_1 i_1=-(-0.1)(0.025)=2.5$ mW. Similarly, in calculating $P_2$, the student writes 0.005 W, which is incorrect. The correct calculation is $P_2=v_2 i_2=(0.2)(0.025)=5$ mW.
	
	\textit{Note:} Llama 3 70B does not recognize the equivalence between 0.005 W and 5 mW after the unit conversion.
\end{itemize}

\subsection{Analysis of Resistive Circuits}
\label{SD22}
\noindent
\textbf{GPT-4o:}
\begin{itemize}
	\item [{\color{red}\ding{171}}]  \texttt{q\_030404\_cs\_37} [Arithmetic]: There are arithmetic errors in your calculations for \( i_1 \) and \( i_2 \).
	
	\begin{itemize}
		\item [--] Correct calculation:
		\[
		i_1=\frac{8}{16+8}(-6)=-2 \text{ A}
		\]
		\[
		i_2=\frac{8}{8+8}(-6)=-3 \text{ A}
		\]
		
		\item [--] Your incorrect calculation:
		\[
		i_1=\frac{1 / 8}{\frac{1}{8}+\frac{1}{8}} \cdot 6 \text{ A}=3 \text{ A}
		\]
		\[
		i_2=\frac{1 / 16}{ \frac{1}{16}+\frac{1}{8}} \cdot 6 \text{ A}=2 \text{ A}
		\]
	\end{itemize}
	
	\textit{Note:} GPT-4o fails to recognize that the student's solution uses a different notation system from the official solution. In this example, the currents \( i_{1} \) and \( i_{2} \) in the student's solution represent \( i_{2} \) and \( i_{1} \), respectively, in the official solution. GPT-4o cannot recognize the exchanged variable notations. To overcome this issue, we can make the variable notations clear in future problem designs.
	
	\item [{\color{red}\ding{171}}]  \texttt{q\_030601\_cs\_27} [Part (c), Final Answer]: The final answer provided, \( i_2 = 0.33\) A, is not accurate and should be \(\frac{1}{3}\) A.
	
	\textit{Note:} The student's final answer \( i_2 = 0.33\) A is acceptable within a rounding error.
	
	\item [{\color{green}\ding{168}}]  \texttt{q\_030601\_cs\_35} [Part (c), Final Answer]:
	
	$i_2=\frac{1}{3} \mathrm{~A}=0.333 \mathrm{~A}$: The value is correct, though it should ideally be presented as $\frac{1}{3} \mathrm{~A}$ to match the given answers more precisely.
	
	\textit{Note:} Different from example \texttt{GPT-4o-q\_030601\_cs\_27}, GPT-4o correctly identifies the value equivalence between \(\frac{1}{3}\) A and 0.333 A. This implies that the performance of GPT-4o is not stable due to its intrinsic randomness when generating responses.
	
	\item [{\color{green}\ding{168}}] \texttt{q\_040202\_cs\_25} [Method]: 
	\begin{itemize}
		\item [--] The student uses Kirchhoff's Voltage Law (KVL) instead of Kirchhoff's Current Law (KCL), which is a valid alternative approach for solving the circuit problem.  
		\item [--] They correctly set up the equations based on the KVL approach, which is an appropriate method.
	\end{itemize}
	
	\textit{Note:} Unlike \texttt{GPT-4o-q\_050405\_cs\_37}, GPT-4o identifies that the method in the student's solution is correct, although the student uses KVL instead of KCL as in the official solution.
	
	\item [{\color{red}\ding{171}}]  \texttt{q\_040205\_cs\_24} [Final Answer]: The final answer in the student's solution is \(0.022 \text{~V}\). While the method and intermediate steps are correct, the final answer should be expressed as \(21.7~ \text{mV}\) (0.0217 V). It appears there is an error in the final conversion or calculation of the voltage difference. 
	
	\textit{Note:} GPT-4o considers the rounding error in the student's solution to be a calculation error.
	
	\item [{\color{red}\ding{171}}]  \texttt{q\_050304\_cs\_14} [Arithmetic]: There are errors in the arithmetic and the process leading to the calculation, primarily in their steps involving the arithmetic for current division:
	\begin{itemize}
		\item [--] For the 9 A source:
		\[
		i_{20A} = 9\text{~A} \cdot \frac{1/35\Omega}{1/10\Omega + 1/35\Omega} \neq 2\text{~A}
		\]
		
		Correct method:
		\[
		i_{20A} = 9\text{~A} \cdot \frac{10\Omega}{10\Omega + 35\Omega} = 9\text{~A} \cdot \frac{10}{45} = 9\text{~A} \cdot \frac{2}{9} = 2\text{~A}
		\]
		
		\item [--] For the 6 A source:
		\[
		i_{20B} = 6\text{~A} \cdot \frac{1/30\Omega}{1/30\Omega + 1/15\Omega} \neq 2\text{~A}
		\]
		
		Correct method:
		\[
		i_{20B} = 6\text{~A} \cdot \frac{15\Omega}{15\Omega + 30\Omega} = 6\text{~A} \cdot \frac{15}{45} = 6\text{~A} \cdot \frac{1}{3} = 2\text{~A}
		\]
	\end{itemize}
	
	\textit{Note:} Although GPT-4o sometimes performs detailed mathematical calculations, the results are not necessarily correct. In this example, we should have $9\text{~A} \cdot \frac{1/35\Omega}{1/10\Omega + 1/35\Omega} = 2\text{~A}$ and $6\text{~A} \cdot \frac{1/30\Omega}{1/30\Omega + 1/15\Omega} = 2\text{~A}$.
	
	\item [{\color{red}\ding{171}}]  \texttt{q\_050405\_cs\_7} [Final Answer]: Unfortunately, your final answer for \( R_t \) is incorrect. The official answer is \( R_t = -\frac{8}{3} \, \Omega \), but you calculated it as \( -\frac{16}{6} \, \Omega \).
	
	\textit{Note:} GPT-4o fails to recognize the equivalence between $-\frac{8}{3}~\Omega$ and $-\frac{16}{6}~\Omega$.
	
	\item [{\color{red}\ding{171}}] \texttt{q\_050405\_cs\_37} [Method]: The student applies KCL to find \(R_t\). However, the method and steps are somewhat unclear and differ from the standard approach usually taught, which directly utilizes the short-circuit current \(i_{sc}\) and open-circuit voltage \(v_{oc}\). This might be confusing.
	
	\textit{Note:} The student's solution uses an alternative approach to find $R_{t}$. Although the method used is different from the approach in the official solution, it is also correct.
\end{itemize}

\noindent
\textbf{Llama 3 70B:}
\begin{itemize}
	\item [{\color{green}\ding{168}}] \texttt{q\_030603\_cs\_5} [Method]: The student uses a correct method to reduce the circuit from the right side by combining series and parallel resistors. However, the student's approach is different from the official solution, which is acceptable as long as the logic is correct.
	
	\textit{Note:} Llama 3 70B correctly identifies an alternative approach as valid.
	
	\item [{\color{green}\ding{168}}] \texttt{q\_040205\_cs\_2} [Final Answer]: The student's final answer, $v = 21.74~\text{mV}$, is correct.
	
	\textit{Note:} The final answer in the official solution is 21.7 mV. Llama 3 70B does not consider the rounding error an arithmetic error.
	
	\item [{\color{green}\ding{168}}] \texttt{q\_040205\_cs\_24} [Final Answer]: The student's final answer, $v = 0.022~\text{V}$, is correct. However, it's worth noting that the problem asks for the answer in millivolts, so the student should have converted the answer to $v = 21.7~\text{mV}$.
	
	\textit{Note:} Llama 3 70B does not consider the rounding error an arithmetic error.
	
	\item [{\color{red}\ding{171}}] \texttt{q\_040303\_ws\_16} [Unit]: The student does not clearly identify the units of the variables throughout the calculation process. For example, the units of the currents $i_1, i_2, i_3$ are not specified.
	
	\textit{Note:} Although the student's solution has deficiencies in all the other four metrics, the student does use the correct units in the solutions. Thus, the above statement about units is incorrect in this case.
	
	\item [{\color{red}\ding{171}}] \texttt{q\_040303\_ws\_28}: 
	
	Completeness: The student's solution is complete, and it answers the question by providing the power supplied by each source.
	
	Final Answers: The student's final answers for $P_{12V}$ and $P_{24V}$ are correct, which are 4.8 W each. However, the student does not explicitly calculate the power supplied by the current source, which is 19.2 W.
	
	\textit{Note:} Although the student does not provide the power supplied by the current source, Llama 3 70B still states that the student's solution is complete.
	
	\item [{\color{green}\ding{168}}] \texttt{q\_040303\_cs\_35} [Method]: The student uses the correct method of applying KCL to node `a' and then solving for the node voltages. However, the student's approach is slightly different from the official solution, which uses a supernode. Both approaches are valid, and the student's method is correct.
	
	\textit{Note:} Llama 3 70B correctly identifies a different approach as valid.
	
	\item [{\color{green}\ding{168}}] \texttt{q\_040706\_ws\_7} [Final Answer]: The student's final answer, $P=-460.8~\text{W}$, is incorrect. The correct answer is $P=460.8~\text{W}$. The student's answer has the correct magnitude but the wrong sign.
	
	\textit{Note:} Llama 3 70B identified the sign error in the student's final answer.
	
	\item [{\color{red}\ding{171}}] \texttt{q\_050201\_cs\_36} [Part (c), Final Answer]: The student's final answer for $i_a$ is incorrect, which is $-\frac{19}{12} \mathrm{A}$ instead of $-1.58 \mathrm{~A}$.
	
	\textit{Note:} Llama 3 70B does not recognize the equivalence between $-\frac{19}{12} \mathrm{A}$ and $-1.58 \mathrm{~A}$.
\end{itemize}

\subsection{The Operational Amplifier}
\label{SD23}
\noindent
\textbf{GPT-4o:}
\begin{itemize}
	\item [{\color{red}\ding{171}}] \texttt{q\_060302\_cs\_2} [Completeness]: The student's solution is incomplete. While it attempts to find the current ($i$) and voltage ($v_o$), it lacks a detailed step-by-step explanation which includes the intermediate steps and proper application of Kirchhoff's Voltage Law (KVL) and Kirchhoff's Current Law (KCL). The discussion concerning $i_2$ and intermediate voltages like $v_a$ is missing.
	
	\textit{Note:} The student's solution is actually complete but misses some step-by-step explanations.
	
	\item [{\color{red}\ding{171}}{\color{green}\ding{168}}] \texttt{q\_060302\_cs\_5} [Arithmetic]:
	There is an arithmetic inconsistency in the presentation. For instance:
	\begin{itemize}
		\item [--] The student jumps from \( 12~= 5000i \) to \( I = \frac{12}{5}~\text{mA} \). The correct \( I \) should be \( I = \frac{12}{5000}~\text{A} \), which equals \( 2.4~\text{mA} \). It seems like an attempt to convert units but the form is incorrect. The numbers are correct, but the portrayal could be cleaner.
	\end{itemize}
	
	\textit{Note:} GPT-4o provides an overly rigorous assessment of the calculation process in the student's solution. Note that GPT-4o correctly identifies \(\frac{12}{5} \text{~mA} = 2.4 \text{~mA}\), which is good.
	
	\item [{\color{green}\ding{168}}] \texttt{q\_060303\_cs\_5} [Method]: 
	The student uses a different approach compared to the official solution. Instead of applying Kirchhoff’s Current Law (KCL) at node \( a \), the student uses the voltage drop across the \( 4~\text{k}\Omega \) resistor to find the current \( i_o \). This method is correct but less direct compared to KCL. The student's reasoning is sound and leads to the correct answer.
	
	\textit{Note:} GPT-4o correctly identifies an alternative approach as correct.
	
	\item [{\color{green}\ding{168}}] \texttt{q\_060303\_cs\_21} [Method]:
	\begin{itemize} 
		\item [--] You correctly identify that the voltages at the input nodes of the op-amp are equal (ideal op-amp assumption). However, the detailed steps for applying KCL at node \(A\) are not shown.
		\item [--] The second equation \( \frac{V_0-V_{A}}{8 \mathrm{k} \Omega}+\frac{V_0}{20 \mathrm{k} \Omega}=0 \) used in your solution is incorrect for this circuit. The inclusion of the \( \frac{V_0}{20 \mathrm{k} \Omega} \) term suggests there might be an incorrect assumption or a misunderstanding about the circuit configuration.
	\end{itemize}
	
	\textit{Note:} The equation \( \frac{V_0 - V_{A}}{8 \,\text{k}\Omega} + \frac{V_0}{20 \,\text{k}\Omega} = 0 \) is indeed incorrect since it implicitly assumes that the output of an operational amplifier always has zero current. Although the final answers in the student's solution are correct, GPT-4o still pointed out this issue. 
	
	\item [{\color{red}\ding{171}}] \texttt{q\_060402\_cs\_7} [Arithmetic]:
	There are no significant arithmetic errors that affect the final result. However, there is a sign issue to note:
	\begin{itemize}
		\item [--] In the equation for \( \text{KCL@} V_b \), you have correctly identified the expression \( \frac{-V_3}{6000}-\frac{V_0}{6000}=0 \), but then you incorrectly simplified it to \( -4 - V_0 = 0 \). The correct simplification should be \( -\frac{4}{6000} - \frac{V_0}{6000} = 0 \) which equates to \( -4 - V_0 = 0 \), so this appears to be more of a transcription error than an arithmetic error.
	\end{itemize}
	\textit{Note:} The response of GPT-4o is redundant since it is completely fine to derive \( -4 - V_0 = 0 \) directly from \( \frac{-V_3}{6000} - \frac{V_0}{6000} = 0 \), given \( V_{3} = 4 \) V.
	
	\item [{\color{red}\ding{171}}{\color{green}\ding{168}}] \texttt{q\_060404\_cs\_37}:
	
	Arithmetic: The student's final expression for \( V_0 \) is:
	
	\[ V_0 = \frac{R_1(V_1 - V_2) - R_3(V_2 - V_1)}{R_2} + V_1 - V_2 \]
	
	Simplifying this,
	
	\[ V_0 = \frac{R_1(V_1 - V_2) + R_3(V_1 - V_2)}{R_2} + V_1 - V_2 \]
	\[ V_0 = \frac{(R_1 + R_3)(V_1 - V_2)}{R_2} + V_1 - V_2 \]
	
	This does not match the official solution:
	
	\[ v_0 = \frac{R_1 + R_2 + R_3}{R_2}(v_1 - v_2) \]
	
	Thus, the student's final answer is incorrect. The steps might have errors that cause the final discrepancy.
	
	Unit: The units are not explicitly addressed in the student's LaTeX representation, but they use standard electrical notations assuming all \( V_x \) terms are in volts and \( R_x \) terms are in ohms. No obvious unit discrepancies are evident.
	
	\textit{Note:} From this example, we can see that GPT-4o has some level of calculation ability. However, it does not fully simplify the original equation for \( V_{0} \) and thus made an incorrect conclusion that the student's final expression does not match the official solution. Additionally, since this problem involves only the calculation of variable notations, we do not need to explicitly list the units. GPT-4o does not consider the omission of units as an error, which is good.
\end{itemize}

\noindent
\textbf{Llama 3 70B:} 
\begin{itemize}
	\item [{\color{red}\ding{171}}] \texttt{q\_060302\_ws\_28} [Unit]: 
	The student's solution lacks clear notation and units. For example, the student uses $V_1$, $V_2$, and $V_0$ without clearly defining what they represent. Additionally, the units of the variables are not consistently identified throughout the calculation process.
	
	\textit{Note:} This is not the case since the student's solution uses the units consistently.
	
	\item [{\color{red}\ding{171}}] \texttt{q\_060303\_cs\_21} [Method]:
	The student uses the correct method, applying KCL at node $A$ and Ohm's law to the $8 \mathrm{k} \Omega$ resistor.
	
	\textit{Note:} The Llama 3 70B model does not correctly identify the incorrect assumption in the methodology.
	
	\item [{\color{green}\ding{168}}] \texttt{q\_060404\_cs\_37}:
	
	Final Answer: The student's final answer is correct, which is $V_0 = \frac{R_1+R_2+R_3}{R_2}(V_1-V_2)$. Although the student's expression looks different from the official solution, it can be simplified to match the official solution.
	
	Units: The student does not explicitly identify the units of the variables throughout the calculation process. However, since the problem does not provide specific values for the resistances and voltages, it is assumed that the student is working with consistent units.
	
	\textit{Note:} Llama 3 70B successfully indicates that the student's final answer is correct, although it is in a different form from the official final answer. However, unlike GPT-4o, Llama 3 70B does not show the detailed mathematical calculation process. Additionally, Llama 3 70B recognizes that it is acceptable to omit the units explicitly since the problem does not provide specific values for the circuit variables.
	
	\item [{\color{red}\ding{171}}] \texttt{q\_060405\_ws\_8} [Unit]: 
	The student does not identify the units of the variables clearly and correctly throughout the calculation process.
	
	\textit{Note:} The solution involves only the notations of variables, not their values, so the units of the variables are implicit. Therefore, it is appropriate that the units are not explicitly defined.
\end{itemize}

\subsection{The Complete Response of Circuits With Energy Storage Elements}
\label{SD24}
\noindent
\textbf{GPT-4o:}
\begin{itemize}
	\item [{\color{red}\ding{171}}{\color{green}\ding{168}}] \texttt{q\_070202\_cs\_2}: 
	
	Correct Final Answer: 
	\begin{itemize}
		\item [--]
		The student's final answer $i(t) = -3 \sin(2t + 30^\circ)$ is not in the correct form. The official solution is given by $i(t) = 3 \cos(2t + 120^\circ)$. There is an error in converting the trigonometric function from sine to cosine form.
	\end{itemize}
	
	Arithmetic Error: 
	\begin{itemize}
		\item [--] The differentiation and multiplication steps are performed correctly. However, there is an error in converting the sine function to the desired cosine function. Specifically, $\sin(x) = \cos(x - 90^\circ)$, thus transforming $-3 \sin(2t + 30^\circ)$ into $3 \cos(2t + 120^\circ)$.
		\item [--] This is an important trigonometric identity that must be correctly applied to reach the problem’s final solution form.
	\end{itemize}
	
	Detailed Feedback: 
	\begin{itemize}
		\item [--] The major issue lies in the final conversion of the trigonometric function. You derived the expression correctly up to $i(t) = -3 \sin(2t + 30^\circ)$. However, to match the standard cosine form, you need to apply the identity $\sin(x) = \cos(x - 90^\circ)$. Hence, $-3 \sin(2t + 30^\circ) = 3 \cos(2t + 30^\circ + 90^\circ) = 3 \cos(2t + 120^\circ)$. 
		
		\item [--] Ensure that you carefully handle trigonometric identities, especially when converting between sine and cosine functions to present your final answer in the required form.
	\end{itemize}
	
	\textit{Note:} In the response about the final answer, GPT-4o indicates that there is an error in converting the trigonometric function from \(-3 \sin(2t + 30^\circ)\) to \(3 \cos(2t + 120^\circ)\). However, GPT-4o also demonstrates how to use the relevant trigonometric identities to correctly convert \(-3 \sin(2t + 30^\circ)\) to \(3 \cos(2t + 120^\circ)\) in the following analysis.
	
	\item [{\color{red}\ding{171}}] \texttt{q\_070202\_cs\_5} [Arithmetic]: The derivative of the current is also incorrect; it should be \( -750e^{-250t} \) not \( -750 e^{\wedge}(-250t) \).
	
	\textit{Note:} GPT-4o is sensitive to the format of mathematical equations.
	
	\item [{\color{green}\ding{168}}] \texttt{q\_070302\_ws\_24} [Final Answer]: The student's final answers have some slight differences, but they are generally correct.
	\begin{itemize}
		\item [--] \( i_c(10 \, \text{ms}) = 8.45 \times 10^{-19} \, \text{A} \) is slightly different from the provided official solution, \( 8.5 \times 10^{-19} \, \text{A} \). This can be due to rounding but should be consistent with significant figures.
		\item [--] \( W(t=10 \, \text{ms}) = 0.125 \, \text{mJ} \) which is the same as \( 1.25 \times 10^{-4} \, \text{J} \).
	\end{itemize}
	
	\textit{Note:} In this example, GPT-4o successfully identifies the rounding error and recognizes the equivalence between two values with different units.
	
	\item [{\color{red}\ding{171}}] \texttt{q\_080301\_cs\_36} [Final Answer]: Your final expressions for \( \tau \) and \( v(t) \) are incorrect. These should match the given solutions, \( \tau = 0.75 \text{ s} \) and \( v(t) = 6 - 2e^{-1.33t} \text{ V} \) for \( t > 0 \). Your exponent in \( v(t) \) should be \( -\frac{t}{\tau} \), where \( \tau = 0.75 \text{ s} \), not \( -\frac{4t}{3} \).
	
	\textit{Note:} The expression $-\frac{t}{\tau}$ is exactly $-\frac{4t}{3}$ when $\tau=0.75$ s.
	
	\item [{\color{red}\ding{171}}] \texttt{q\_080401\_cs\_5} [Arithmetic]: There are a few arithmetic inaccuracies:
	\begin{itemize}
		\item [--] You calculated \(v(1.5) = 5.0027 \) V instead of the correct $v(1.5) = 5$ V. 
		\item [--] You used $v(1.5) = 5.0027$ V for the subsequent calculations, which caused a slight deviation in your final expression for $t > 1.5$ s.
	\end{itemize}
	
	\textit{Note:} GPT-4o considers the small discrepancy between 5 V and 5.0027 V an arithmetic inaccuracy rather than a rounding error.
	
	\item [{\color{green}\ding{168}}] \texttt{q\_080401\_cs\_11} [Arithmetic]: The initial calculation of \( V(1.5^-)=5 + 5 e^{-5 \cdot 1.5} \) seems approximately correct. However, rounding to 5.003 V is unnecessary and should ideally be kept as 5 V based on typical conventions.
	
	\textit{Note:} GPT-4o considers the small discrepancy between 5 V and 5.003 V to be a rounding error.
	
	\item [{\color{red}\ding{171}}{\color{green}\ding{168}}] \texttt{q\_090201\_cs\_7} [Final Answer]: No, the student's final answers have some issues. The final solution derived is:
	
	\[ 10^8 V_s = \frac{d^2 V}{d T^2} + 3000 \frac{d V}{d T} + 1.02 \times 10^8 V \]
	
	However, the correct final answer is:
	
	\[ 1 \times 10^8 v_{\mathrm{s}}=1.02 \times 10^8 v+3000 \frac{\mathrm{d}v}{\mathrm{d}t}+\frac{\mathrm{d}^2 v}{\mathrm{d}t^2} \]
	
	Notably, the order and coefficients of terms in the student’s final equation match, but there are discrepancies in the variables used ( \(V\) vs. \(v\) and \(T\) vs. \(t\)). 
	
	\textit{Note:} Although GPT-4o identifies that the coefficients of the terms in the student's final equation match, it still indicates that the student's final answer is incorrect due to discrepancies in the variables used. 
	
	\item [{\color{red}\ding{171}}] \texttt{q\_090201\_cs\_19} [Final Answer]: Correctness of Final Answer:
	\begin{itemize}
		\item [--] The final equation \( 1 \times 10^8 v_s = \frac{d^2 v}{dt^2} + 3000 \frac{dv}{dt} + 1.02 \times 10^8 v_a \) the student reaches is incorrect primarily because of coefficient and term sign errors throughout the derivation.
		
		\item [--] This divergence from the correct answer \( 1 \times 10^8 v_{\mathrm{s}} = 1.02 \times 10^8 v + 3000 \frac{\mathrm{d}v}{\mathrm{d}t} + \frac{\mathrm{d}^2 v}{\mathrm{d}t^2} \) indicates a need to revisit algebraic and differential handling within this setup.
	\end{itemize}
	
	\textit{Note:} GPT-4o fails to recognize the equivalence between two differential equations in which the terms are arranged in different orders.
	
	\item [{\color{red}\ding{171}}{\color{green}\ding{168}}] \texttt{q\_090303\_cs\_5} [Final Answer]: The student's final answers for the roots of the characteristic equation are $s = -3031.260$ and $s = -98969.740$. These values are close to the correct answers of $s = -3031$ and $s = -98969$, but the precision given is slightly different. The correct roots are indeed theoretically rounded to these values, but the more precise values provided by the student suggest a deeper discrepancy in the calculation steps.
	
	\textit{Note:} GPT-4o successfully recognizes the small deviations in the student's final answers as rounding errors. However, it incorrectly suggests a deeper discrepancy in the calculation steps.
	
	\item [{\color{red}\ding{171}}] \texttt{q\_090303\_cs\_7} [Arithmetic]: Specifically, right before solving for the characteristic equation, you have inaccurately arrived at \(1.02 \times 10^5 s \) instead of the correct \(102000 s\).
	
	\textit{Note:} GPT-4o does not recognize the equivalence between \(1.02 \times 10^5\) s and 102,000 s.
	
	\item [{\color{red}\ding{171}}] \texttt{q\_090303\_cs\_23} [Final Answer]: 
	\begin{itemize}
		\item [--] The final numerical solutions for \( s_1 \) and \( s_2 \) are incorrect. The official correct answers are \( s_1 = -3031 \), \( s_2 = -98969 \). 
		\item [--] The student has \( S_1 = -3031.26 \) and \( S_2 = -98968.74 \). Despite small deviations, these values are incorrect due to errors in prior steps. Additionally, the discrepancy here would not be considered rounding errors.
	\end{itemize}
	
	\textit{Note:} GPT-4o does not recognize the small deviations in the student's final answers as rounding errors.
\end{itemize}

\noindent
\textbf{Llama 3 70B:}
\begin{itemize}
	\item [{\color{red}\ding{171}}] \texttt{q\_070202\_cs\_2} [Final Answer]: The student's final answer is incorrect. The correct final answer is $i(t) = 3 \cos (2t + 120^{\circ}) ~\text{A}$, but the student's answer is $i(t)=-3 \sin (2 t+30)$.
	
	\textit{Note:} Llama 3 70B does not identify the equivalence between $3 \cos (2t + 120^{\circ})$ and $-3 \sin (2 t+30)$.
	
	\item [{\color{red}\ding{171}}] \texttt{q\_070204\_cs\_5} [Final Answer]: 
	\begin{itemize}
		\item [--] The unit of voltage is incorrect in some places. For example, in the first interval, the voltage should be $-10^{-3}~\text{V}$, not $-1~\text{mV}$.
		\item [--] The coefficient of $t$ in the second interval is incorrect. It should be $2\times 10^6$, not $2,000,000$.
		\item [--] The coefficient of $t$ in the third interval is incorrect. It should be $-10^6$, not $-1,000,000$.
	\end{itemize}
	
	\textit{Note:} In this example, Llama 3 70B does not identify the equivalence between different forms of a value, indicating its unstable performance in this aspect. 
	
	\item [{\color{red}\ding{171}}{\color{green}\ding{168}}] \texttt{q\_070302\_cs\_2}:
	
	Arithmetic: There is a small arithmetic error in the calculation of $i(10 \mathrm{~ms})$. The correct value is $8.5 \times 10^{-19} \mathrm{~A}$, but the student gets $8.45 \times 10^{-19} \mathrm{~A}$. This is a minor error and does not affect the overall correctness of the solution.
	
	Unit: The units for all variables are identified clearly and correctly throughout the calculation process, except for the energy at $t=10$ ms, where the unit should be J instead of mJ.
	
	\textit{Note:} Llama 3 70B seems to recognize the rounding errors in the final answer since it states that the minor error does not affect the overall correctness of the solution. However, it is overly rigorous about the units, as both J and mJ are acceptable for the final answer. 
	
	\item [{\color{red}\ding{171}}{\color{green}\ding{168}}] \texttt{q\_070302\_ws\_24} [Final Answer]: The student's final answers are mostly correct. They correctly find the current $i_c(0)=0.2$ A and $i_c(10\text{ ms})=8.45\times 10^{-19}$ A, which matches the official solution. However, there is a minor issue with the energy calculation. The student finds the energy at $t=10$ ms to be $0.125$ mJ, which is correct, but they should have written it in scientific notation as $1.25\times 10^{-4}$ J to match the official solution.
	
	\textit{Note:} Llama 3 70B successfully recognizes the equivalence between $0.125$ mJ and $1.25\times 10^{-4}$ J. However, it provides an overly rigorous assessment, as the final answer does not have to be written in scientific notation.
	
	\item [{\color{red}\ding{171}}] \texttt{q\_070504\_cs\_5} [Final Answer]: The student's final answer is almost correct. The correct final answer is $v(t) = 15 \sin(500t + 60^{\circ})$ V, while the student's answer is $v(t) = 15 \sin(500t + \frac{\pi}{3})$. The phase angle in the student's answer is incorrect. 
	
	\textit{Note:} Llama 3 70B does not identify the equivalence between the degree $60^{\circ}$ and radian $\frac{\pi}{3}$. 
	
	\item [{\color{green}\ding{168}}] \texttt{q\_080301\_cs\_35} [Arithmetic]: 
	There is a small arithmetic error in the calculation of the time constant $\tau$. The student writes $\tau = 3 \times 250 \times 10^{-3} = 0.75~\text{s}$, which is correct. However, in the final answer, the student uses $e^{-\frac{t}{2}}$ instead of $e^{-\frac{t}{\tau}} = e^{-1.33t}$. This is a minor mistake, but it does not affect the overall correctness of the solution.
	
	\textit{Note:} Llama 3 70B seems to recognize that there is a typo in $e^{-\frac{t}{2}}$, which should correctly be expressed as $e^{-\frac{t}{\tau}}$.
	
	\item [{\color{red}\ding{171}}] \texttt{q\_080301\_cs\_36} [Final Answer]: 
	The student's final answer is incorrect. The correct final answer is $v(t) = 6 - 2e^{-1.33t}$ V for $t > 0$, whereas the student's answer is $v(t) = 6 - 2e^{-4t/3}$ V.
	
	\textit{Note:} Llama 3 70B fails to identify the equivalence between $v(t) = 6 - 2e^{-1.33t}$ V and $v(t) = 6 - 2e^{-4t/3}$ V.
	
	\item [{\color{red}\ding{171}}] \texttt{q\_080401\_cs\_5} [Arithmetic]: 
	There are some arithmetic errors in the student's solution. For example, $V(1.5)=5.0027$ V is an approximation, but the student should have used the exact value $V(1.5)=5$ V. This error propagates to the second part of the solution, resulting in an incorrect coefficient in the exponential term.
	
	\textit{Note:} Llama 3 70B identifies the rounding error in the student's solution as an arithmetic error. 
	
	\item [{\color{green}\ding{168}}] \texttt{q\_090201\_cs\_7} [Final Answer]: The student's final answer is correct: $10^8 V_s=\frac{d^2 V}{d T^2}+3000 \frac{d V}{d T}+1.02 \times 10^8 V$, which matches the official solution.
	
	\textit{Note:} The differential equation provided in the official solution is \( 1 \times 10^8 v_{\mathrm{s}} = 1.02 \times 10^8 v + 3000 \frac{\mathrm{d}v}{\mathrm{d}t} + \frac{\mathrm{d}^2 v}{\mathrm{d}t^2} \). Llama 3 70B successfully identified the equivalence between the two differential equations.
	
	\item [{\color{green}\ding{168}}] \texttt{q\_090201\_cs\_19} [Final Answer]: The student's final answer is correct, although it is written in a slightly different form than the official solution. The student's answer is: $1 \times 10^8 v_s=\frac{d^2 v}{d t^2}+3000\frac{d v}{d t}+1.02 \times 10^8 v_a$, which is equivalent to the official solution: $1 \times 10^8 v_s=1.02 \times 10^8 v+3000 \frac{d v}{d t}+\frac{d^2 v}{d t^2}$.
	
	\textit{Note:} Llama 3 70B successfully identified the equivalence between the two differential equations.
	
	\item [{\color{green}\ding{168}}] \texttt{q\_090303\_cs\_23} [Final Answer]: 
	The student's final answers, $s_1=-3031.26$ and $s_2=-98968.74$, are correct, considering the rounding errors. The official solution has $s_1=-3031$ and $s_2=-98969$, which are very close to the student's answers.
	
	\textit{Note:} Llama 3 70B correctly identified the rounding errors in the student's final answers. 
\end{itemize}

\subsection{Sinusoidal Steady-State Analysis}
\label{SD25}
\noindent
\textbf{GPT-4o:}
\begin{itemize}
	\item [{\color{red}\ding{171}}] \texttt{q\_100302\_ws\_11} [Completeness]: Unfortunately, the student's solution is not complete as it fails to arrive at the correct final answer and some intermediate steps are either skipped or improperly detailed.
	
	\textit{Note:} The student's solution is complete as it answers the question posed by the problem.
	
	\item [{\color{red}\ding{171}}] \texttt{q\_100302\_cs\_21}: 
	
	Final Answer: The student's final answer is:
	
	\[
	V(t)=\sqrt{74} \cos \left(8 t-9.462^{\circ}\right)
	\]
	
	While the correct answer should be:
	
	\[
	v(t)=8.602 \cos \left(8 t-9.462^{\circ}\right) ~\text{V}
	\]
	
	It can be observed that the student's amplitude calculation is incorrect, and hence the final answer deviates from the correct one. 
	
	Units: Throughout the calculation, it is important to note that voltages are correctly identified with the units `V' and specified parameters like angles are correctly in degrees.
	
	\textit{Note:} Since \(\sqrt{74} \approx 8.602\), the student's final answer matches the official solution up to a rounding error. Additionally, the student did not consistently use units for all variables.
	
	\item [{\color{red}\ding{171}}] \texttt{q\_100406\_cs\_23} [Completeness]: The student's solution is incomplete. While the student performs some calculations and finds the voltage across the \(40~\Omega \) resistor, there are crucial mistakes, and the final answer does not match the required format. The student fails to clearly convert the phasors back to the time domain correctly and correct the unit errors.
	
	\textit{Note:} The student's solution is complete since it answers the question posed by the problem.
	
	\item [{\color{red}\ding{171}}] \texttt{q\_100601\_cs\_24} [Final Answer]:
	The student's final answer \( 14.69 \angle 5.55^{\circ} \) is close but not correct. The correct answer is \( 14.67 \angle 5.6^{\circ} \).
	
	\textit{Note:} GPT-4o does not recognize the value discrepancies in the student's final answer as rounding errors.
	
	\item [{\color{red}\ding{171}}{\color{green}\ding{168}}] \texttt{q\_100604\_cs\_24} [Final Answer]:
	The final answers given by the student for the currents are almost correct but slightly off:
	\begin{itemize}
		\item [--] \( \mathbf{I}_1 = 0.794 \angle 111^{\circ} \)
		\item [--] \( \mathbf{I}_2 = 0.789 \angle -61.7^{\circ} \) (official answer is 0.790, the angle matches though)
		\item [--] \( \mathbf{I}_3 = 0.228 \angle 176.3^{\circ} \) (official answer is 0.229 \(\angle 176^{\circ}\))
	\end{itemize}
	These minor discrepancies suggest either small arithmetic errors or possibly rounding issues. They should recalculate these to ensure accuracy.
	
	\textit{Note:} GPT-4o identifies the small discrepancies in the final answers as possible rounding errors. However, these values can be considered correct and do not need to be recalculated.
	
	\item [{\color{red}\ding{171}}] \texttt{q\_100701\_ws\_5} [Unit]:
	\begin{itemize}
		\item [--] Consistently use proper notation for complex numbers and phasors.
		\item [--] Provide final results in correct units—Volts for $\mathbf{V}_{\mathrm{oc}}$ and Ohms for $\mathbf{Z}_{\mathrm{t}}$. 
	\end{itemize}
	
	\textit{Note:} The student, in effect, did not consistently use units for all variables. 
	
	\item [{\color{red}\ding{171}}] \texttt{q\_100801\_cs\_25} [Final Answer]:
	Your final expression for $i(t)$ is:
	\[ i(t) = 0.036 \cos\left(4000 t - 26.57^{\circ}\right) + 0.036 \cos (2500 t - 153.43) \]
	This is incorrect. The official solution's final answer is:
	\[ i(t) = 35.8 \cos \left(2500 t - 153.44^{\circ}\right) + 35.8 \cos \left(4000 t - 26.56^{\circ}\right) \text{mA} \]
	
	\textit{Note:} GPT-4o does not recognize the equivalence between the student's final expression and the official final expression.
\end{itemize}

\noindent
\textbf{Llama 3 70B:}
\begin{itemize}
	\item [{\color{red}\ding{171}}] \texttt{q\_100302\_cs\_21} [Unit]: The student correctly identifies the units of the variables throughout the calculation process.
	
	\textit{Note:} The student did not consistently use units throughout the calculation process.
	
	\item [{\color{green}\ding{168}}] \texttt{q\_100401\_ws\_8} [Final Answer]: $Z_C = -j2.4~\Omega$ (student's answer is correct, but it should be written in the form of $-j2.4~\Omega$ instead of $\frac{2.4}{j}$).
	
	\textit{Note:} Llama 3 70B effectively identified that \(-j2.4 = \frac{2.4}{j}\).
	
	\item [{\color{green}\ding{168}}] \texttt{q\_100406\_ws\_2} [Final Answer]: The student's final answer is correct: $v(t)=0.9662 \cos \left(4000 t-140^{\circ}\right) ~\text{V}$. Although the student's solution is written in terms of current ($I_{40\Omega}$) instead of voltage ($V$), the calculation is correct, and the final answer can be easily obtained by multiplying the current by the resistance ($R=40\Omega$).
	
	\textit{Note:} The student's solution only reaches $I_{40\Omega}=24.15\angle-139.89^{\circ} \text{~mA}$, but not the required voltage $v(t)=0.9662 \cos \left(4000 t-140^{\circ}\right) ~\text{V}$. Llama 3 70B correctly identifies that the final answer for $v(t)$ can be easily obtained by multiplying the current by the resistance ($R=40\Omega$), which is good.
	
	\item [{\color{green}\ding{168}}] \texttt{q\_100601\_cs\_24}: 
	
	Final Answer: The student's final answer for $V_o$ is close to the correct answer, but not exactly equal. The student's answer is $14.69 \angle 5.55^{\circ}$, while the correct answer is $14.67 \angle 5.6^{\circ}$. This is a minor error, but it is still an error.
	
	Arithmetic: There are no obvious arithmetic errors in the student's solution. However, there might be some minor calculation errors that led to the slight difference in the final answer.
	
	\textit{Note:} Llama 3 70B does not consider the slight difference in the final answer, which is due to the rounding error during the calculation process, as a significant calculation error.
	
	\item [{\color{red}\ding{171}}] \texttt{q\_100701\_ws\_21} [Unit]: The units of all variables are identified clearly and correctly throughout the calculation process.
	
	\textit{Note:} The student did not consistently use units throughout the calculation process.
	
	\item [{\color{red}\ding{171}}] \texttt{q\_100801\_ws\_5} [Final Answer]: The student's final answer is incorrect. The correct final answer is $i(t)=i_1(t)+i_2(t)=[35.8 \cos (2500 t-153.44^{\circ})+35.8 \cos (4000 t-26.56^{\circ})] ~\text{mA}$. The student's answer has different values for the amplitude and phase angles. 
	
	\textit{Note:} The correct final answer is \( i_{\mathrm{o}}(t) = \frac{2}{25\sqrt{5}} \cos (2500 t + 26.57^{\circ}) + \frac{2}{25\sqrt{5}} \cos (4000 t - 26.57^{\circ}) \) A. Thus, the student's answer has the correct values for the amplitudes because \( \frac{2}{25\sqrt{5}} \text{ A} \approx 35.8 \text{ mA} \).
	
	\item [{\color{green}\ding{168}}] \texttt{q\_100801\_ws\_21} [Final Answer]: The student's final answer is incorrect. The student's answer is in amperes (A), but it should be in milliamperes (mA). Additionally, the student's answer has a typo in the cosine arguments, where $2500 \omega$ should be $2500 t$ and $4000 \omega$ should be $4000 t$.
	
	\textit{Note:} Llama 3 70B explicitly pointed out the important typo in the student's solution.
	
	\item [{\color{red}\ding{171}}] \texttt{q\_100801\_cs\_25} [Final Answer]: The student's final answer is almost correct, but there is a unit error. The student's answer is in amperes (A), whereas the correct unit is milliamperes (mA). The correct final answer is: $i(t) = 35.8 \cos (2500 t - 153.44^{\circ}) + 35.8 \cos (4000 t - 26.56^{\circ}) \text{ mA}$.
	
	\textit{Note:} Llama 3 70B is overly rigorous about the units. The student's final answer of $i(t)=0.036 \cos \left(4000 t-26.57^{\circ}\right)+0.036 \cos (2500 t-153.43) \text{~A}$ is also fully correct.
	
	\item [{\color{red}\ding{171}}] \texttt{q\_100806\_cs\_5} [Unit]: The units of all variables are identified clearly and correctly throughout the calculation process.
	
	\textit{Note:} The student did not consistently use units throughout the calculation process.
\end{itemize}

\subsection{Frequency Response}
\label{SD26}
\noindent
\textbf{GPT-4o:}

\begin{itemize}
	\item [{\color{green}\ding{168}}] \texttt{q\_130203\_cs\_5} [Final Answer]: 
	Your answer $H(\omega) = \frac{3}{5} \times \frac{1}{1 + j\omega\frac{4}{5}}$ translates to \(H(\omega) = \frac{0.6}{1 + j\omega0.8}\), which is algebraically equivalent to the correct final answer \(H(\omega) = \frac{0.6}{1 + j(0.8)\omega}\). While your final answer is mathematically accurate, it should ideally match the official solution's format for clarity.
	
	\textit{Note:} GPT-4o successfully identified the equivalence between $\mathbf{H}(\omega)=\frac{0.6}{1+j(0.8) \omega}$ and $\mathbf{H}(\omega)=\frac{3}{5}\cdot\frac{1}{1+j\omega\frac{4}{5}}$.
	
	\item [{\color{green}\ding{168}}] \texttt{q\_130209\_ws\_24} [Final Answer]: 
	The student's expression for the gain magnitude is incorrect.
	\begin{itemize}
		\item [--] Student's expression: \( |H(\omega)| = 5 \frac{\sqrt{1 + 16 \times 10^{-4} \omega}}{\sqrt{1 + 1 \times 10^{-2} \omega}} \)
		\item [--] Correct expression: \( |H(\omega)| = 5 \frac{\sqrt{1 + \frac{\omega^2}{625}}}{\sqrt{1 + \frac{\omega^2}{100}}} \)
		\item [--] Error: The student incorrectly simplifies the squared \(\omega\) terms.
	\end{itemize}
	
	\textit{Note:} GPT-4o correctly indicates that there are incorrect simplifications for the squared $\omega$ terms.
\end{itemize}

\noindent
\textbf{Llama 3 70B:}
\begin{itemize}
	\item [{\color{red}\ding{171}}] \texttt{q\_130201\_cs\_10}: 
	
	Final Answer: The student's final answer is not correct. The student's answer is $\frac{10}{50+200 \omega j}$, whereas the correct answer is $\frac{0.2}{1+j 4 \omega}$. The student's answer has a different numerator and denominator compared to the official solution.
	
	Unit: The units of the variables are not clearly identified throughout the calculation process. For example, the student uses $\omega c j$ in the expression for $V_0(\omega)$, but the units of $C$ are not specified. Similarly, the units of $\omega$ are not specified. In the official solution, it is clear that $C$ is in Farads and $\omega$ is in radians per second.
	
	\textit{Note:} Similar to example \texttt{Llama-3-70B-q\_130203\_cs\_5}, Llama 3 70B does not recognize the equivalence between $\frac{10}{50+200 \omega j}$ and $\frac{0.2}{1+j 4 \omega}$ in this example. In addition, it is acceptable that the units of \(\omega\) are not specified during the calculation process.
	
	\item [{\color{red}\ding{171}}] \texttt{q\_130202\_cs\_3} [Unit]: 
	The units of all variables are identified clearly and correctly throughout the calculation process. The student correctly uses k$\Omega$ for resistances and $\mu$F for capacitance.
	
	\textit{Note:} This is a hallucination of Llama 3 70B, as the student never mentioned the units k$\Omega$ and $\mu$F in their solution.
	
	\item [{\color{green}\ding{168}}] \texttt{q\_130202\_cs\_9} [Method]: 
	The student uses a different method than the official solution, but it is still a correct approach. Instead of directly finding the transfer function $H(\omega)$, the student uses the impedance and nodal analysis to find the ratio of $V_o$ to $V_i$, which is equivalent to $H(\omega)$.
	
	\textit{Note:} Llama 3 70B correctly identifies an alternative approach as correct. 
	
	\item [{\color{red}\ding{171}}] \texttt{q\_130203\_cs\_5}:
	
	Final Answer: The student's final answer is not entirely correct. While they have derived the correct expression for $\mathbf{H}(\omega)$, they haven't simplified it to the final form. The correct final answer is $\mathbf{H}(\omega)=\frac{0.6}{1+j(0.8) \omega}$, whereas the student's answer is $\mathbf{H}(\omega)=(3/5)*1/(1+jw4/5)$, which is close but not identical.
	
	Unit: The student has not explicitly identified the units of the variables throughout the calculation process. While it's not a major error, it's essential to clearly identify the units of variables to avoid confusion.
	
	\textit{Note:} In terms of the final answer, Llama 3 70B does not find the equivalence between \(\mathbf{H}(\omega)=\frac{0.6}{1+j(0.8) \omega}\) and \(\mathbf{H}(\omega)=\frac{3}{5}\cdot\frac{1}{1+j\omega\frac{4}{5}}\). Regarding units, their inclusion is sometimes unnecessary during the calculation process of a network function.
	
	\item [{\color{red}\ding{171}}{\color{green}\ding{168}}] \texttt{q\_130208\_cs\_11} [Unit]: 
	The student does not explicitly identify the units of the variables throughout the calculation process. However, based on the context of the problem, it can be inferred that $R_1$ and $R_2$ are in k$\Omega$, C is in $\mu$F, and $\omega$ is in rad/s. It would be better if the student explicitly stated the units of each variable.
	
	\textit{Note:} Llama 3 70B seems to have some ability to infer the units of circuit variables. However, it is acceptable that the student did not explicitly state the units of the variables when calculating a network function with variable notations.
	
	\item [{\color{red}\ding{171}}] \texttt{q\_130209\_ws\_24} [Final Answer]: 
	The student's final answers are not entirely correct.
	\begin{itemize}
		\item [--] The transfer function $H(\omega)$ is close, but the student has a typo in the numerator, writing $j0.04w$ instead of $j\omega/25$. Similarly, in the denominator, it should be $j\omega/10$ instead of $j0.1w$.
		\item [--] The gain $|H(\omega)|$ is close, but the student has a typo in the numerator, writing $16 \times 10^{-4}w$ instead of $\omega^2/625$. Similarly, in the denominator, it should be $\omega^2/100$ instead of $1 \times 10^{-2}w$.
		\item [--] The phase shift $\angle H(\omega)$ is close, but the student has a typo in the second term, writing $+\tan^{-1}(0.1w)$ instead of $-\tan^{-1}(\omega/10)$.
	\end{itemize}
	\textit{Note:} Llama 3 70B does not recognize the equivalence between different representations of an expression.
\end{itemize}

\end{appendices}

\bibliography{bibliography}

\section*{Statements and Declarations}
The authors appreciate the support provided by the School of Electrical and Computer Engineering and the College of Engineering at the Georgia Institute of Technology. The authors would also like to acknowledge the assistance of ChatGPT in polishing the language of this paper.

\subsection*{Competing Interests}
The authors have no relevant financial or non-financial interests to disclose. 

\subsection*{Author Contributions}
The conceptualization and methodology design were carried out by Liangliang Chen and Ying Zhang. The assessment evaluation metrics were designed by Liangliang Chen, Jacqueline Rohde, and Ying Zhang.  Data collection and dataset construction were performed by Liangliang Chen, Zhihao Qin, Yiming Guo, and Ying Zhang. The classification of LLM responses, i.e., data analysis, was conducted by Liangliang Chen. The initial draft of the manuscript was prepared by Liangliang Chen and subsequently reviewed and edited by Jacqueline Rohde and Ying Zhang. Project management and funding acquisition were handled by Ying Zhang. All authors have read and approved the final manuscript. 

\subsection*{Data Availability}
The dataset used in this study was collected from an undergraduate-level circuit analysis course during the Fall 2021 semester at the Georgia Institute of Technology. Students' solutions will not be shared with the community, as the students enrolled in the Fall 2021 course did not sign a consent form. The official reference solutions are available in the solution manual for the textbook: \textit{Svoboda, J.A., Dorf, R.C.: Introduction to Electric Circuits (9th Edition). John Wiley \& Sons, (2013)}, which is subject to copyright. The LLM response data used in our evaluation of the performance of different LLMs is available upon request. Additionally, we are actively gathering more student homework data from various courses, including circuit analysis, and plan to release datasets from students who have provided explicit consent. 

\end{document}